\documentclass{article}
\usepackage{arxiv}
\usepackage[utf8]{inputenc}
\usepackage[T1]{fontenc}
\usepackage[english]{babel}
\usepackage{amsmath}
\usepackage{amsthm}
\usepackage{amssymb,amsfonts}
\usepackage{mathrsfs}
\usepackage{bbm}
\usepackage{graphicx}
\usepackage{enumitem}
\usepackage{xcolor}
\RequirePackage[colorlinks,citecolor=blue,urlcolor=blue]{hyperref}
\usepackage[normalem]{ulem}
\usepackage{pgf,tikz}

\usepackage{dsfont}
\usepackage{fancyhdr}
\usepackage{mathtools}
\usepackage[ruled,vlined]{algorithm2e}
\usepackage{caption}
\usepackage{subcaption}
\usepackage{booktabs} 
\usepackage{parcolumns}
\usepackage{breqn}

\usepackage{parskip,url,multirow,arydshln,afterpage}
\usepackage[breakable,theorems]{tcolorbox}
\usepackage{enumerate}
\usepackage[title]{appendix}

\usepackage{pgf,tikz}
\usetikzlibrary{calc}
\usetikzlibrary{arrows,automata,shapes,positioning}
\usetikzlibrary{arrows.meta,fit,backgrounds}

\usepackage{wasysym}

\newcommand{\margarida}[1]{\colorbox{teal}{\color{white}   \textsf{\textbf{Margarida}}} \textcolor{teal}{#1}}

\newcommand{\Stwo}{$\Sigma_2^p$}
\newcommand{\Sth}{$\Sigma_3^p$}
\newcommand{\MCN}{MCN}

\newcommand{\ie}{\emph{i.e.}}

\newcommand{\boxxx}[1]
 {\begin{center}\fbox{\begin{minipage}{12.50cm}#1\smallskip\end{minipage}}\end{center}}
 
\newtheorem{thm}{Theorem}[section]
\newtheorem{lem}[thm]{Lemma}

\newtheorem{prop}[thm]{Property}

\title{On the completeness of several fortification-interdiction games in the polynomial hierarchy}

\author
{
Alberto Boggio Tomasaz\\
Universit\`a degli Studi di Milano\\
Dipartimento di Informatica, Milano, Italy\\
\and
Margarida Carvalho\\
CIRRELT and Département d’Informatique et de Recherche Opérationnelle\\
Universit\'e de Montr\'eal, Montr\'eal, QC, Canada\\
\and
Roberto Cordone\\
Universit\`a degli Studi di Milano\\
Dipartimento di Informatica, Milano, Italy\\
\and
Pierre Hosteins\\
Universit\`a degli Studi di Torino, \\
Dipartimento di Informatica, Torino, Italy \\
e-mail: \texttt{hosteins@di.unito.it} \\
}

\date{\today}

\begin{document}
\maketitle

\begin{abstract}
Fortification-interdiction games are tri-level adversarial games where two opponents act in succession to protect, disrupt and simply use an infrastructure for a specific purpose. Many such games have been formulated and tackled in the literature through specific algorithmic methods, however very few investigations exist on the completeness of such fortification problems in order to locate them rigorously in the polynomial hierarchy. We clarify the completeness status of several well-known fortification problems, such as the Tri-level Interdiction Knapsack Problem with unit fortification and attack weights, the Max-flow Interdiction Problem and Shortest Path Interdiction Problem with Fortification, the Multi-level Critical Node Problem with unit weights, as well as a well-studied electric grid defence planning problem. For all of these problems, we prove their completeness either for the \Stwo{} or the \Sth{} class of the polynomial hierarchy\footnote{The demonstration of the \Stwo-completeness of the Bi-level Interdiction Knapsack Problem provided in Section~\ref{sec:BIKP} has also been accepted for publication in the proceedings of the \emph{International Conference on Optimization and Decision Science ODS2024}.}. We also prove that the Multi-level Fortification-Interdiction Knapsack Problem with an arbitrary number of protection and interdiction rounds and unit fortification and attack weights is complete for any level of the polynomial hierarchy, therefore providing a useful basis for further attempts at proving the completeness of protection-interdiction games at any level of said hierarchy. 
\end{abstract}


\section{Introduction}
\label{sec:Intro}

The field of interdiction games has known a substantial amount of research in the last two decades. Such problems, also called \emph{Stackelberg games}, represent the sequential action of two opposite players who seek to optimise the same objective function but in opposite directions. By nature such problems are bi-level, with a \emph{follower} agent reacting to the decision of a \emph{leader} agent. In their basic version, the follower, also named \emph{defender}, tries to perform a certain action in the best possible way
after the leader, or \emph{attacker}, has invested a limited amount of resources to disable or disrupt some element of the problem. For example, the defender aims to maximise a flow, find a shortest path between an origin and a destination node, select a set of items given some capacity constraint, etc... whereas the attacker disables items to pack, edges or nodes of a network, etc... Such problems have become very popular, as they allow us to understand the weak points of, e.g., critical infrastructures and to assess the robustness of such infrastructures to worst-case failures or malevolent attacks. They have a wide range of possible applications, from the analysis of electric power grids \cite{Salmeron2009} or water distribution networks \cite{Wu2021} to telecommunications~\cite{ARULSELVAN20092193}, nuclear smuggling~\cite{Sullivan2014b}, drug network dismantling~\cite{malaviya}, public healthcare~\cite{assimakopoulos}, and so on.

The most classical interdiction problems are based on the attack of network elements to disrupt classic graph problems, such as maximising a quantity of flow (Maximum Flow Interdiction Problem or MFIP~\cite{Wood1993}) or minimising the length of the shortest path between an origin and a destination node (Shortest Path Interdiction Problem or SPIP~\cite{Israeli2002}). While these problems usually consider attacks on network arcs, other interdiction problems focus more on identifying the most vital nodes whose removal has the largest impact on network connectivity, such as the Critical Node Problem (CNP), see~\cite{LALOU201892} for a survey of the literature on the CNP. Even though those problems are bi-level in nature, they are based on a lower level problem which is polynomially solvable. Hence, these problems belong to the class NP and are usually NP-complete in the general case~\cite{Addis2013,Israeli2002,Wood1993}. More recently, harder interdiction problems have been considered, where the lower level is NP-hard and which are therefore not necessarily located in NP. This is the case of an interdiction version of the Knapsack Problem (KP), called Bi-level Interdiction Knapsack Problem (BIKP), where a leader, subject to an attack budget, can interdict a subset of items which cannot be packed by the follower when maximising the profit of the basic KP~\cite{Caprara2016,Croce2020A,DeNegre2011,Weninger2023}. A version of this problem where the lower level KP weights are equal to the item profits is proved to be \Stwo-complete, along with other bi-level KPs, in~\cite{Caprara2014}. Other interdiction problems with an NP-hard lower level have been shown to be \Stwo-complete, such as the Maximum Clique Interdiction Problem~\cite{Furini2019} and the Facility Location Interdiction Problem~\cite{Frohlich2021}.

Since the interdiction games often have the aim of identifying the most vital elements of the problem at hand, a natural extension of such problems is to add another layer of decision in order to protect some of the problem elements against a possible attack. For example, an extension of the CNP (which aims at fragmenting a graph into separate maximally connected components) is proposed in~\cite{Barbosa2021} to add a certain number of edges in the graph prior to the nodes attack, in order to increase the robustness of the graph. The most common intervention studied to protect against interdiction is called \emph{fortification}, i.e., rendering a subset of elements \emph{immune} to any attack. The resulting problems have been labelled as Defender-Attacker-Defender (DAD) problems~\cite{Brown2006}. Virtually any existing interdiction problem can be extended to a fortification version of the same problem. Examples of such fortification problems, numerically tackled in the literature with several algorithmic approaches, are the Shortest Path Interdiction Problem with Fortification (SPIPF)~\cite{Lozano2017,LEITNER20231026}, the Tri-level Interdiction Knapsack Problem (TIKP)~\cite{LEITNER20231026}, the Capacitated Lot Sizing Interdiction Problem with Fortification (CLSIPF)~\cite{Lozano2017} or the Travelling Salesman Problem with Interdiction and Fortification (TSPIF)~\cite{Lozano2017b}. Focusing more on the application side, some specific fortification problems have been investigated, such as the fortification version of an electric power grid interdiction problem, which we will call Tri-level Electric Power Grid Fortification Problem (TEPGFP)~\cite{Chen2023,Fakhry2022,Wu2017}, the fortification of water distribution networks~\cite{Wu2021} or the defence of critical infrastructures in homeland security~\cite{Brown2006}. A variant of the CNP with fortification has also been considered, where the attacker can infect a subset of the graph nodes with a virus which will automatically spread through the graph edges, while a defender first vaccinates a subset of nodes against the virus and later protects a subset of nodes through quarantine after the attacker has acted. This problem is called Multi-level Critical Node Problem (MCNP) and has been introduced in~\cite{Baggio2021}.

Even though many works have tackled tri-level problems with fortification, surprisingly few of them address the computational complexity of such problems. Given that tri-level linear programming is in general \Stwo-complete~\cite{Johannes2011NewCO}, tri-level fortification problems will in general not be in class NP, which implies that they will require advanced, sophisticated decomposition methods in order to be solved~\cite{Baggio2021,Wu2017}. In particular, no compact single level reformulation exists for such problems. Known complexity results include the \Sth-completeness of the TIKP with equal weights and profits for the lower level KP and the \Sth-completeness of the MCNP with arbitrary weights (vaccination, infection and protection weights as well as profits following the terminology introduced in the previous paragraph) 
over trees~\cite{NABLI2022122}. A few results also exist for related problems in robust programming, such as the \Sth-completeness of several two-stage adjustable and robust recoverable optimisation problems with \emph{discrete} budgeted uncertainty~\cite{Goerigk2024}. We observe that, in general, fortification problems are tackled with unitary fortification and attack costs, meaning that the fortification and interdiction budget constraints are actually cardinality constraints on the subsets of fortified and interdicted objects, probably for reasons of simplicity in designing the problem instances to tackle numerically. Hence, complexity results for such ``unitary'' versions are desirable.

\paragraph{Paper contributions:}
We clarify the computational complexity of several fortification problems, either directly tackled in the literature or serving as a basis for proving the completeness of other fortification problems in the future. More specifically, we prove:
\begin{itemize}
    \item The completeness for any level of the polynomial hierarchy of the Multi-level Interdiction Knapsack Problem with an arbitrary number of  fortification and interdiction rounds, with unit fortification and attack costs (therefore refining the results of~\cite{Caprara2014,NABLI2022122}).
    \item The \Stwo-completeness of the SPIPF and TEPGFP with unit fortification costs, and of the MFIP with Fortification (MFIPF) with unit costs for both fortification and attack.
    \item The \Sth-completeness of the MCNP with unit weights (at any level) and profits, refining the result of~\cite{NABLI2022122}.
\end{itemize}

\paragraph{Paper Organisation} In Section~\ref{sec:Knapsack}, we will tackle the interdiction version of the KP and show that the bi-level and tri-level versions are, respectively, \Stwo{} and \Sth-complete with unit attack and fortification costs, closing the section with a generalisation of our proof to any level of the polynomial hierarchy. In Section~\ref{sec:MaxFlow}, we move to the fortification version of the MFIP and prove that it is \Stwo-complete, still with unit attack and fortification costs. In Section~\ref{sec:ShortestPath} we prove the \Stwo-completeness of the fortification version of the SPIP with unit fortification costs. Moving to an interdiction-fortification problem in which the attacker deletes nodes, we will demonstrate the \Sth-completeness of the MCNP with only unit weights and profits in Section~\ref{sec:MCNP}, therefore demonstrating completeness on a more general set of instances w.r.t. to the work of \cite{NABLI2022122}. Finally, we will clarify the status of the TEPGFP in Appendix~\ref{sec:PowerGrid} by establishing its \Stwo-complete nature and provide our conclusions.

Given the wide variety of problems tackled, where possible we keep the notation consistent with the one used in the original papers. However, we will also try and make it more uniform when necessary to avoid confusion, in particular when dealing with multi-level versions of the same problem.

\section{The Multi-level Interdiction Knapsack Problem}
\label{sec:Knapsack}

In this section we will tackle the interdiction version of the Knapsack Problem, as presented in Section~\ref{sec:Intro}. As already stated earlier, the computational complexity of the bi-level and tri-level versions of this problem have been investigated in the literature, for specific versions where each item has an arbitrary fortification and attack cost, plus a packing weight for the lower level (classic KP) which is equal to the item profit~\cite{Caprara2014,NABLI2022122}. We will consider the slightly more general version where the lower level packing weights are not equal to the profits. However, we will concentrate on versions where the fortification and attack costs are unitary. This could be useful in future endeavours on other fortification problems where such costs are also unitary, as it will allow to reduce the interdiction knapsack problems to them. Unfortunately, the reductions proposed in the previous works of \cite{Caprara2014,NABLI2022122} are highly dependent on the presence of attack and fortification costs, which are used to differentiate between the knapsack items that are mapped to the different levels of the problems used for the reduction. It is therefore impossible to use those reductions as proposed in those works when the attack and fortification costs are unitary. 
We stress that our results are also interesting from a purely theoretical perspective, since in the single level KP, computational complexity tends to be strongly affected when a set of weights/profits becomes unitary (as the problem goes from being weakly NP-hard to polynomial). It is therefore a legitimate concern to establish the consequences of adopting unit costs for the higher decision levels in interdiction versions of the KP. In this section, we will establish that even with unitary fortification and attack costs, the BIKP and TIKP are respectively complete for the second and third level of the polynomial hierarchy. We will close this section by sketching how to extend the proof of completeness for any number of fortification and interdiction rounds, showing that the problem is complete for any level of the polynomial hierarchy.

\subsection{The Bi-level Interdiction Knapsack Problem with unit attack costs}
\label{sec:BIKP}

Our definition of the decision version of the Bi-level Interdiction Knapsack Problem with unit attack costs is the following. The indices used for sets and parameters refer to the decision level (1 for the defender, 2 for the attacker), for consistency with the multi-level generalisation in Section \ref{SubSec:MIKP}. We also adopt the convention that, when dealing with the decision version of the optimisation problems considered in this work, we drop the word \emph{Problem} at the end of their name, or the \emph{P} at the end of their acronym.
\vspace{0.3cm}
\boxxx{
\textsc{\textbf{Unitary Bilevel Interdiction Knapsack (UBIK)}}: \\
{\sc instance}: A finite set of items $I$ such that each $i \in I$ has a positive integer weight $w_i$ and a positive integer profit $p_i$, a positive integer capacity $W$ and a positive integer profit goal $\bar{K}$ for the follower, a positive integer budget $B_2$ for the leader.\\
{\sc question}: Is there a subset $I_2\subseteq I$ of items for the attacker to forbid, with cardinality $|I_2| \leq B_2$, such that every subset $I_1\subseteq I\setminus I_2$ with $\sum_{i \in I_1} w_i \leq W$ that the defender can select has a total profit $\sum_{i \in I_1} p_i < \bar{K}$?
}
\vspace{0.3cm}
In order to prove that the above problem is complete for the second level of the polynomial hierarchy, we will use the following problem, known to be \Stwo-complete \cite{Wrathall1976}:
\vspace{0.3cm}
\boxxx{
\textbf{\textsc{2-Alternating Quantified Satisfiability}} ($\mathcal B_2 \cap \overline{3CNF}$):\\
{\sc instance}: Disjoint non-empty sets of Boolean variables $X$ and $Y$, and a Boolean expression $E$ over $U = X \cup Y$ in conjunctive normal form with at most 3 literals in each clause $c \in C$. \\
{\sc question}: Is there a truth assignment for $X$ such that for all truth assignments of $Y$, $E$ is never satisfied?
}
\vspace{0.3cm}

We will then introduce a reduction from $\mathcal B_2 \cap \overline{3CNF}$ to UBIK which draws inspiration from the reduction from 3-SAT to Subset Sum presented in~\cite[Theorem 34.15]{IntrAlg2009}. The basic idea of this reduction is to build weights, profits, capacity and profit goal of UBIK encoded in base 10, with digits associated with variables and clauses of $\mathcal B_2 \cap \overline{3CNF}$, in such a way that any comparison between a sum of integer values (weights or profits) and a reference integer value (capacity or profit goal) can be performed digit by digit, without carries from lower digits to higher ones. 
\begin{itemize}
    \item  For each variable $u \in U$, we create two items $i_u$ and $i_{\bar{u}}$, one for each possible truth assignment of $u$. We define $I_X = \lbrace i_x,i_{\bar x}: x \in X \rbrace$ and $I_{Y} = \lbrace i_y,i_{\bar{y}}: y \in Y \rbrace$. For variables $x\in X$, we introduce a third item $j_x$, which will serve to force the leader to choose one and only one item between $i_x$ and $i_{\bar{x}}$. In order to ensure a sensible mapping of the variable assignment between $Y$ and $I_Y$, we also introduce a fourth item $j_x^\prime$ for $x\in X$ to help saturate the budget of the follower for $X$-related digits (leaving therefore only the $Y$-related digits of the budget $W$ available for selecting items from $I_Y$ in $I_1$).
    \item For each clause $c \in C$, we create two items $i^1_c$, $i^{2}_c$. We designate by $I_C$ the set of items associated with $C$.
    \item  
    We now introduce weights and profits for all the created items through a set of digits, each multiplied by a different power of 10, as previously discussed. This is examplified in Table~\ref{B2_ubik} where these digits are displayed in the different columns of the table and form a (large) number when put right next to each other. The least significant $\vert C \vert$ digits are labelled by the clauses, the next $\vert Y \vert$ digits by the variables $Y$, the next 2$\vert X \vert$ digits by the variables $X$, two for each variable $x\in X$. Finally, the most significant positions form a set starting at position $M = \vert C\vert +\vert Y \vert +2\vert X \vert +1$. For the sake of simplicity, with a slight abuse of notation, we will sometimes refer to the integer value encoded in these positions as a ``digit''.
        \begin{itemize}
            \item For each $y \in Y$, the two corresponding items $i_y$ and $i_{\bar{y}}$ have weights and profits ($w_{i_y}$, $p_{i_y}$, $w_{i_{\bar{y}}}$ and $p_{i_{\bar{y}}}$) with digit 1 in the position labelled by the variable $y$ and 0 in the positions labelled by other variables.

            If literal $y$ appears in clause $c \in C$, then $p_{i_y}$ and $w_{i_y}$ have digit 1 in the position labelled as $c$, and 0 otherwise. Similarly, if literal $\neg y$ appears in clause $c \in C$,  $p_{i_{\bar{y}}}$ and $w_{i_{\bar{y}}}$ have digit 1 in the position labelled by $c$, and 0 otherwise. 

            In the small example detailed in Table~\ref{B2_ubik}, for example, variable $c$ is the first of the two variables in $Y$ and appears in the third (and last) clause, therefore we have $w_{i_c}=p_{i_c}=10,001$.
            
            \item For each $x \in X$, the two corresponding items $i_x$ and $i_{\bar{x}}$ have weights $w_{i_x}$ and $w_{i_{\bar{x}}}$ and profits $p_{i_x}$ and $p_{i_{\bar{x}}}$ with digit 1 in the higher of the two positions labelled by variable $x$ and 0 in the lower one and in the positions labelled by other variables.

            If literal $x$ appears in clause $c \in C$, then $p_{i_x}$ and $w_{i_x}$ have digit 1 in the position labelled as $c$, and 0 otherwise. Similarly, if literal $\neg x$ appears in clause $c \in C$, $p_{i_{\bar{x}}}$ and $w_{i_{\bar{x}}}$ have digit 1 in the position labelled by $c$, and 0 otherwise. The weights and profits have digit 0 in all remaining positions. As the first variable in $X$, present in the 1st and 3rd clause, variable $a$ from the example of Table~\ref{B2_ubik} provides the weight $w_{i_a}=100,000,101$ and profit $p_{i_a}=1,100,000,101$ which differ by the leftmost digit.

            The item $j_x$ corresponding to $x$ has weight and profit with digit 2 in the higher position labelled as $x$ (and 0 elsewhere).

            The item $j_x^\prime$ corresponding to $x$ has weight $w_{j_x^\prime}$ with digit 1 in the higher position labelled as $x$ and profit $p_{j_x}$ with digit 1 in the lower one
            (and 0 elsewhere).

            Using again the example of Table~\ref{B2_ubik} with variable $a$, we have for item $j_a$ that $w_{j_a}=p_{j_a}=200,000,000$ with a digit only in the left column associated to $a$ and for item $j_a^\prime$ we have $w_{j_a^\prime}=100,000,000$ and $p_{j_a^\prime}=10,000,000$ where the profit has a digit only in the right column associated to $a$.
            
            \item For each $c \in C$, the first item has weight and profit with digit 1 in the position labelled as $c$ and 0 elsewhere, and the second item has weight and profit with digit 2 in the position labelled as $c$ and 0 elsewhere. For the first clause out of three in Table~\ref{B2_ubik}, we have that $w_{i_{c_1}^1}=p_{i_{c_1}^1}=100$ while $w_{i_{c_2}^1}=p_{i_{c_2}^1}=200$.
            \item The attack budget $B_2$ is equal to $|X|$.
            \item The capacity $W$ has digit 1 in all positions labelled as variables in $Y$, 2 in all higher positions labelled as variables in $X$, 4 in all positions labelled as clauses in $C$, and digit 0 elsewhere. Hence, $I_1$ can contain any item from $I_X$ (as long as not interdicted).
            \item The profit goal $\bar{K}$ has digit 4 for all positions with labels in $C$ and 1 for all positions with labels in $U$. The positions starting at $M$ encode the integer value $|X|$. 
        \end{itemize}
\end{itemize}
See Figure~\ref{B2_ubik} for an illustration of the proposed reduction. 
The reduction is polynomial because the profit goal has $\vert C\vert + \vert Y \vert + 2\vert X \vert + \lceil \log_{10} \vert X \vert \rceil$ decimal digits and the other numbers are smaller.

\begin{figure}
\footnotesize
    \centering
    \resizebox{0.5\textwidth}{!}{
    \begin{tabular}{ll|c|cccc:cc|ccc}
    $I$           &                    &       & \multicolumn{4}{c:}{$X$} & \multicolumn{2}{c|}{$Y$} & \multicolumn{3}{c}{$C$} \\  
                  &                    & $M$   & \multicolumn{2}{c}{$a$} & \multicolumn{2}{c}{$b$} & $c$ & $d$ & $c_1$ & $c_2$ & $c_3$ \\ \midrule 

    $i_a$         & $w_{i_a}$          & 0     & 1 & 0   & 0 & 0   & 0   & 0   &  1    &  0    & 1 \\
                  & $p_{i_a}$          & 1     & 1 & 0   & 0 & 0   & 0   & 0   &  1    &  0    & 1 \\
    $i_{\bar{a}}$ & $w_{i_{\bar{a}}}$  & 0     & 1 & 0   & 0 & 0   & 0   & 0   &  0    &  1    & 0 \\
                  & $p_{i_{\bar{a}}}$  & 1     & 1 & 0   & 0 & 0   & 0   & 0   &  0    &  1    & 0 \\
    $j_a$         & $w_{j_a}$          & 0     & 2 & 0   & 0 & 0   & 0   & 0   &  0    &  0    & 0 \\
                  & $p_{j_a}$          & 0     & 2 & 0   & 0 & 0   & 0   & 0   &  0    &  0    & 0 \\
    $j_a^\prime$  & $w_{j_a^\prime}$   & 0     & 1 & 0   & 0 & 0   & 0   & 0   &  0    &  0    & 0 \\
                  & $p_{j_a^\prime}$   & 0     & 0 & 1   & 0 & 0   & 0   & 0   &  0    &  0    & 0 \\
    $i_b$         & $w_{i_b}$          & 0     & 0 & 0   & 1 & 0   & 0   & 0   &  1    &  0    & 1 \\
                  & $p_{i_b}$          & 1     & 0 & 0   & 1 & 0   & 0   & 0   &  1    &  0    & 1 \\
    $i_{\bar{b}}$ & $w_{i_{\bar{b}}}$  & 0     & 0 & 0   & 1 & 0   & 0   & 0   &  0    &  1    & 0 \\
                  & $p_{i_{\bar{b}}}$  & 1     & 0 & 0   & 1 & 0   & 0   & 0   &  0    &  1    & 0 \\
    $j_b$         & $w_{j_b}$          & 0     & 0 & 0   & 2 & 0   & 0   & 0   &  0    &  0    & 0 \\
                  & $p_{j_b}$          & 0     & 0 & 0   & 2 & 0   & 0   & 0   &  0    &  0    & 0 \\
    $j_b^\prime$  & $w_{j_b^\prime}$   & 0     & 0 & 0   & 1 & 0   & 0   & 0   &  0    &  0    & 0 \\
                  & $p_{j_b^\prime}$   & 0     & 0 & 0   & 0 & 1   & 0   & 0   &  0    &  0    & 0 \\
    $i_c$         & $w_{i_c}$          & 0     & 0 & 0   & 0 & 0   & 1   & 0   &  0    &  0    & 1 \\ 
                  & $p_{i_c}$          & 0     & 0 & 0   & 0 & 0   & 1   & 0   &  0    &  0    & 1 \\ 
    $i_{\bar{c}}$ & $w_{i_{\bar{c}}}$  & 0     & 0 & 0   & 0 & 0   & 1   & 0   &  1    &  0    & 0 \\
                  & $p_{i_{\bar{c}}}$  & 0     & 0 & 0   & 0 & 0   & 1   & 0   &  1    &  0    & 0 \\
    $i_d$         & $w_{i_d}$          & 0     & 0 & 0   & 0 & 0   & 0   & 1   &  0    &  1    & 0 \\
                  & $p_{i_d}$          & 0     & 0 & 0   & 0 & 0   & 0   & 1   &  0    &  1    & 0 \\
    $i_{\bar{d}}$ & $w_{i_{\bar{d}}}$  & 0     & 0 & 0   & 0 & 0   & 0   & 1   &  0    &  0    & 0 \\
                  & $p_{i_{\bar{d}}}$  & 0     & 0 & 0   & 0 & 0   & 0   & 1   &  0    &  0    & 0 \\
    $i^1_{c_1}$   & $w_{i^1_{c_1}}$    & 0     & 0 & 0   & 0 & 0   & 0   & 0   & 1     & 0     & 0 \\
                  & $p_{i^1_{c_1}}$    & 0     & 0 & 0   & 0 & 0   & 0   & 0   & 1     & 0     & 0 \\
    $i^2_{c_1}$   & $w_{i^2_{c_1}}$    & 0     & 0 & 0   & 0 & 0   & 0   & 0   & 2     & 0     & 0 \\
                  & $p_{i^2_{c_1}}$    & 0     & 0 & 0   & 0 & 0   & 0   & 0   & 2     & 0     & 0 \\
    $i^1_{c_2}$   & $w_{i^1_{c_2}}$    & 0     & 0 & 0   & 0 & 0   & 0   & 0   & 0     & 1     & 0 \\
                  & $p_{i^1_{c_2}}$    & 0     & 0 & 0   & 0 & 0   & 0   & 0   & 0     & 1     & 0 \\
    $i^2_{c_2}$   & $w_{i^2_{c_2}}$    & 0     & 0 & 0   & 0 & 0   & 0   & 0   & 0     & 2     & 0 \\
                  & $p_{i^2_{c_2}}$    & 0     & 0 & 0   & 0 & 0   & 0   & 0   & 0     & 2     & 0 \\
    $i^1_{c_3}$   & $w_{i^1_{c_3}}$    & 0     & 0 & 0   & 0 & 0   & 0   & 0   & 0     & 0     & 1 \\
                  & $p_{i^1_{c_3}}$    & 0     & 0 & 0   & 0 & 0   & 0   & 0   & 0     & 0     & 1 \\
    $i^2_{c_3}$   & $w_{i^2_{c_3}}$    & 0     & 0 & 0   & 0 & 0   & 0   & 0   & 0     & 0     & 2 \\
                  & $p_{i^2_{c_3}}$    & 0     & 0 & 0   & 0 & 0   & 0   & 0   & 0     & 0     & 2 \\ \midrule
                  & $W$                & 0     & 2 & 0   & 2 & 0   & 1   & 1   & 4     & 4     & 4  \\
                  & $\bar{K}$          & 2     & 1 & 1   & 1 & 1   & 1   & 1   & 4     & 4     & 4  
    \end{tabular}
    }
    \caption{Example of construction of UBIK from an instance $\mathcal B_2 \cap \overline{3CNF}$ with $E =(a \lor b \lor \neg c) \land (\neg a \lor \neg b \lor d) \land (a \lor b \lor c) $, where $X = \{a, b \}$, $Y = \{c,d\}$ and the clauses are labelled from left to right. Given the digits provided in the different columns, we have, e.g., for item $i_a$: $w_{i_a}=100,000,101$ and $p_{i_a}=1,100,000,101$.}
    \label{B2_ubik}
\end{figure}

Analysing an instance from the above reduction, we can infer the following properties:
\begin{prop}\label{prop1_UBIK}
If the leader does not interdict either $i_x$ or $i_{\bar x}$ for each $x\in X$, the follower can reach the profit goal $\bar K$.
\end{prop}
\begin{proof}
First of all, the attack budget forbids to interdict more than $|X|$ items. If the leader interdicts strictly less than $|X|$ items of $I_X$, the capacity of the follower allows to select all the uninterdicted items from $I_X$, reaching the profit goal $\bar K$. Hence, only those solutions where the leader interdicts exactly $|X|$ items from $I_X$ (and no item $j_x$ or $j_x^\prime$ for $x\in X$) will be of interest for the rest of the proof. For such solutions, the follower has enough capacity to select all uninterdicted items from $I_X$ and must select them (any other possibility leading to a failure to reach the profit goal $\bar K$). Suppose that the two items $i_x$ and $i_{\bar{x}}$ associated with the most significant digits whose label is $x\in X$ are taken simultaneously in $I_2$. In this case, the follower can select item $j_x$, as well as all remaining uninterdicted items of $I_X$, achieving a profit of $|X|10^{M}+2\times10^{|C|+|Y|+2|X|}>\bar K$. In the example of Table~\ref{B2_ubik}, consider, e.g., that the leader interdicts both $i_a$ and $i_{\bar a}$: the follower has enough capacity to select $i_b$, $i_{\bar b}$ and also $j_a$, which provides a profit of $2,200,000,000>\bar K$. Suppose instead that the leader selects neither $i_x$ nor $i_{\bar x}$: since, as already stated, the follower has enough budget to select all remaining $I_X$ items, it is possible for $I_1$ to contain both $i_x$ and $i_{\bar x}$ and the profit goal $\bar K$ is automatically achieved, as the positions from $M$ on have value $|X|$ and the highest digit associated with $x$ is 2.
The leader must therefore interdict one item between $i_x$ and $i_{\bar x}$. Then, the follower can include item $j_x^\prime$ in $I_1$ in order to reach digit 1 in the lower position associated to $x$ (while not enough residual capacity is available to select item $j_x$), and must do that 
in order to be able to reach profit $\bar K$. Consequently, the capacity associated to both digits associated to $x$ is saturated.

This reasoning can then be recursively continued until we reach the lowest $X$-related position, ensuring the above property.
\end{proof}

\begin{prop}\label{prop2_UBIK}
If the leader interdicts either $i_x$ or $i_{\bar x}$ for each $x\in X$, 
in order to reach the profit goal $\bar K$ the follower must necessarily select the remaining $|X|$ uninterdicted items of $I_X$, all $j'$ items and either $i_y$ or $i_{\bar y}$ for $y\in Y$.
\end{prop}
\begin{proof}
Suppose the leader does interdict exactly one item between $i_x$ and $i_{\bar x}$ for each $x\in X$. The follower has enough capacity to select all remaining uninterdicted $I_X$ items and must do so, otherwise it is impossible to reach value $|X|$ at position $M$. Following the reasoning in the proof of Property~\ref{prop1_UBIK}, the follower must also select all $j_x^\prime$ items, in order to reach digit 1 in all the lowest digits associated to each $x\in X$, therefore saturating the capacity associated to all $X$ variables and forbidding to select any $j$ item. In order to be able to reach the profit goal, the follower must also select a subset of items from $I_Y$:
\begin{itemize}
    \item Consider the most significant digit whose label is in $Y$. Its two associated items cannot be taken simultaneously in $I_1$ as it would violate the weight capacity $W$. Taking none of them would forbid to achieve a profit of $\bar{K}$. Therefore, exactly one of these items must be taken, saturating the associated capacity.
    \item Consider the second most significant digit whose label is in $Y$. Its two associated items cannot be taken simultaneously, since that would result in a violation of the weight capacity $W$, and cannot be both neglected if the profit goal must be reached. Hence, as before, $I_1$ will include exactly one of the items associated with the second most significant digit in $Y$.
    \item The reasoning above propagates until the least significant digit labelled in $Y$. 
\end{itemize}
We conclude that $I_1$ must include either $i_y$ or $i_{\bar{y}}$ for all $y \in Y$.
\end{proof}

\begin{thm}
UBIK is \Stwo-complete.
\end{thm}
\begin{proof}
The statement of UBIK is of the form $\exists I_2 \ \ \forall I_1 \ \ Q(I_2,I_1)$ where $Q$ can be tested in polynomial time, directly implying that it is in \Stwo.

As mentioned above, we use a reduction from $\mathcal B_2 \cap \overline{3CNF}$ to prove the \Stwo-hardness of UBIK.

Let $\mathcal B_2 \cap \overline{3CNF}$ be a \emph{Yes} instance. We form a solution to UBIK by packing in $I_2$ the items $i_x$ such that $x \in X$ is 1 and the items $i_{\bar{x}}$, otherwise. Clearly, the cardinality of $I_2$ is equal to $|X|$.
By Property~\ref{prop2_UBIK}, in order to try and reach the profit goal, the follower must include in $I_1$ all $|X|$ items associated with $X$ not packed in $I_2$, all $j_x^\prime$ items for $x\in X$ and exactly one of the items $i_y$ or $i_{\bar{y}}$ for $y \in Y$.
$I_1$ can be completed by items from $I_C$. However, since the $\mathcal B_2 \cap \overline{3CNF}$ instance is a \emph{Yes} instance, there is at least a clause for which no item from $I_X$ and $I_Y$ with a non-zero value in the associated digit can be selected. For the digit corresponding to that clause, the prize of set $I_1$ cannot reach value 4 using only items of $I_C$ and respecting the budget $W$. Therefore, UBIK is a \emph{Yes} instance.

Next, suppose that UBIK is a \emph{Yes} instance. According to Properties~\ref{prop1_UBIK} and \ref{prop2_UBIK}, $I_2$ must contain exactly one of the items $i_x$ and $i_{\bar{x}}$ for $x \in X$ and $I_1$ will contain exactly one item between $i_y$ and $i_{\bar{y}}$ for each $y\in Y$, otherwise the follower cannot achieve profit $\bar K$. However, since UBIK is a \emph{Yes} instance, no choice of the follower allows to reach a value of 4 for all digits associated to $C$. This means that for at least one digit associated to $C$, none of the items of $X$ and $Y$ with a digit 1 in said position can be selected. Assign value 1 to $x \in X$ such that $i_{\bar{x}} \in I_2$, and 0 otherwise. For any set $I_1$ of the follower, assign 1 to $y \in Y$ if $i_{\bar{y}}\in I_1$ and 0 if $i_y\in I_1$. Since, by hypothesis, UBIK is a \emph{Yes} instance, there is no assignment of the $Y$ variables such that $E$ is satisfied in the corresponding SAT instance and the $\mathcal B_2 \cap \overline{3CNF}$ instance is a \emph{Yes} instance.
\end{proof}

Although the above proof follows a basis similar to the proof of \Sth-completeness of the Tri-level Interdiction Knapsack Problem in \cite{NABLI2022122}, it required the introduction of new features in order to deal with the symmetry between the different items from the point of view of the attacker, given the equal value of all the interdiction costs. This is the meaning of the additional $M$ column (to favour the choice of the $I_X$ items by the attacker) and of the $j_x$ and $j_x^\prime$ items (to force the attacker to interdict either $i_x$ or $i_{\bar x}$).

\subsection{The Tri-level Interdiction Knapsack Problem with unit attack and fortification costs}

We now demonstrate the completeness of the TIKP with unit attack and fortification costs for the third level of the polynomial hierarchy. This version of the TIKP is defined as:
\vspace{0.3cm}
\boxxx{
\textsc{\textbf{Unitary Trilevel Interdiction Knapsack (UTIK)}}: \\
{\sc instance}: A set of items $I$ such that each $i \in I$ has a positive integer weight $w_i$ and a positive integer profit $p_i$, a positive integer capacity $W$ and a positive integer profit goal $\bar{K}$, two positive integer budgets $B_2$ and $B_3$ for attack and fortification, respectively.\\
{\sc question}: Is there a subset $I_3\subseteq I$ of items, with $|I_3| \leq B_3$, such that for every subset $I_2\subseteq I\setminus I_3$, with $|I_2| \leq B_2,$ there is a subset $I_1 \subseteq I \setminus I_2$, with $\sum_{i \in I_1} w_i \leq W$, such that $\sum_{i \in I_1} p_i \geq \bar{K}$ holds?
}
\vspace{0.3cm}
In order to prove the \Sth-completeness of the above problem, we will use the following problem, known to be \Sth-complete \cite{Stockmeyer1976,Wrathall1976}:
\vspace{0.3cm}
\boxxx{
\textbf{\textsc{3-Alternating Quantified Satisfiability}} ($\mathcal B_3 \cap 3CNF$):\\
{\sc instance}: Disjoint non-empty sets of variables $X$, $Y$ and $Z$, and a Boolean expression $E$ over $U = X \cup Y \cup Z$ in conjunctive normal form with at most 3 literals in each clause $c \in C$. \\
{\sc question}: Is there a 0-1 assignment for $X$ such that for all 0-1 assignments of $Y$ there is a 0-1 assignment of $Z$ such that $E$ is satisfied?
}
We will then use a reduction from $\mathcal B_3 \cap 3CNF$ similar to the one used to demonstrate the \Stwo-completeness of UBIK:
\begin{itemize}
    \item For each variable $u \in U$, we create two items $i_u$ and $i_{\bar{u}}$, one for each possible 0-1 assignment of $u$. We define $I_X = \lbrace i_x,i_{\bar x}: x \in X \rbrace$, $I_{Y} = \lbrace i_y,i_{\bar{y}}: y \in Y \rbrace$ and $I_{Z} = \lbrace i_z,i_{\bar{z}}: z \in Z \rbrace$. For variables of the intermediate decision level $y\in Y$, we introduce a third item $j_y$, which will serve to force the attacker to choose one and only one item between $i_y$ and $i_{\bar{y}}$, as well as a fourth item $j_y^\prime$ which will serve to force the defender to choose one and only one item between $i_z$ and $i_{\bar{z}}$ for each $z\in Z$.
    \item For each clause $c \in C$, we create two items $i^1_c$, $i^{2}_c$. We designate by $I_C$ the set of items associated with $C$.
    \item Weights, profits, maximum capacities, maximum profit and goal are positive integer numbers expressed in base 10.
    Each digit position is labelled by a variable or a clause, except the largest ones: the least significant $\vert C \vert$ positions are labelled by the clauses, the following $\vert Z \vert$ positions by the variables $Z$, the next $2\vert Y \vert$ positions by the variables $Y$ (two consecutive digits for each variable), then the next $\vert X \vert$ positions are labelled  by the variables $X$.
    Finally, the most significant digits form a set starting at position $M = \vert C\vert +\vert Y \vert +2\vert X \vert +1$. 
        \begin{itemize}
            \item For each $u \in X\cup Z$, the two corresponding items $i_u$ and $i_{\bar{u}}$ have weights and profits $w_{i_u}$, $p_{i_u}$, $w_{i_{\bar{u}}}$ and $p_{i_{\bar{u}}}$ as described next. The weights and profits have digit 1 in the position labelled by the variable $u$ and 0 in the positions labelled by the other variables.
            
            If literal $u$ appears in clause $c \in C$, then $p_{i_u}$ and $w_{i_u}$ have digit 1 in the position labelled as $c$, and 0 otherwise. Similarly, if literal $\neg u$ appears in clause $c \in C$,  $p_{i_{\bar{u}}}$ and $w_{i_{\bar{u}}}$ have digit 1 in the position labelled by $c$, and 0 otherwise.
            
            Finally, if $u\in X$, $p_{i_u}$ and $p_{i_{\bar{u}}}$ have value $|Y|+1$ in the positions from $M$ on.
            \item For each variable $y \in Y$, the weights and profits of the two corresponding items $i_y$ and $i_{\bar{y}}$ have digit 1 in the higher of the two positions labelled by $y$ and 0 in the lower one and in the positions labelled by other variables. 
            
            The profits $p_{i_y}$ and $p_{i_{\bar{y}}}$ also have digit 1 in position $M$. If literal $y$ appears in clause $c \in C$, then $p_{i_y}$ and $w_{i_y}$ have digit 1 in the position labelled as $c$, and 0 otherwise. Similarly, if literal $\neg x$ appears in clause $c \in C$, $p_{i_{\bar{y}}}$ and $w_{i_{\bar{y}}}$ have digit 1 in the position labelled by $c$, and 0 otherwise.
            
            The item $j_y$ corresponding to $y$ has a weight and profit with digit 2 in the highest position labelled as $y$.

            Finally, the item $j_y^\prime$ has a profit with digit 1 in the lower position associated to $y$ and a weight with a digit 1 in the higher position associated to $y$, with a 0 in all remaining positions.
            \item For each $c \in C$, the first item has weight and profit with digit 1 in the position labelled as $c$ and 0 elsewhere, and the second item has weight and profit with digit 2 in the position labelled as $c$ and 0 elsewhere.
            \item The fortification budget $B_3$ is set equal to $|X|$ and the attack budget $B_2$ to $|Y|$.
            \item The capacity $W$ has 1s for all digits with labels in $Z\cup X$, 2s for all digits associated with the higher position of each label in $Y$, 4s for all digits with labels in $C$, and 0s elsewhere. Hence, $I_3$ can contain any $I_Y$ item (as long as not interdicted).
            \item The profit goal $\bar{K}$ has 1s for all digits with labels in $U$, 4s for all digits with labels in $C$ and the decimal encoding of $(|Y|+1)|X|+|Y|$ in the digits starting at position $M$.
        \end{itemize}
\end{itemize}

See Figure~\ref{B3_utik} for an illustration of the proposed reduction on an instance with the same overall set of variables $U$ as before (distributed now over three subsets) and the same Boolean expression as in the example of Figure~\ref{B2_ubik}. The reduction is polynomial because the profit goal takes $\vert C\vert + \vert Z \vert + 2 \vert Y \vert + \vert X \vert + \lceil \log_{10} [ (|Y|+1)|X|+|Y|) ] \rceil$ decimal digits and the other numbers are smaller.

\begin{figure}
\footnotesize
    \centering
    \resizebox{0.5\textwidth}{!}{
    \begin{tabular}{ll|c|c:cc:cc|ccc}
    $I$           &                    &       & $X$ & \multicolumn{2}{c}{$Y$} & \multicolumn{2}{:c}{$Z$} & \multicolumn{3}{c}{$C$} \\  
                  &                    & $M$ & $a$ & \multicolumn{2}{c}{$b$} & $c$ & $d$ & $c_1$ & $c_2$ & $c_3$ \\ \midrule 
    $i_a$         & $w_{i_a}$          & 0   & 1   & 0 & 0   & 0   & 0   &  1    &  0    & 1 \\
                  & $p_{i_a}$          & 2   & 1   & 0 & 0   & 0   & 0   &  1    &  0    & 1 \\
    $i_{\bar{a}}$ & $w_{i_{\bar{a}}}$  & 0   & 1   & 0 & 0   & 0   & 0   &  0    &  1    & 0 \\
                  & $p_{i_{\bar{a}}}$  & 2   & 1   & 0 & 0   & 0   & 0   &  0    &  1    & 0 \\
    $i_b$         & $w_{i_b}$          & 0   & 0   & 1 & 0   & 0   & 0   &  1    &  0    & 1 \\
                  & $p_{i_b}$          & 1   & 0   & 1 & 0   & 0   & 0   &  1    &  0    & 1 \\
    $i_{\bar{b}}$ & $w_{i_{\bar{b}}}$  & 0   & 0   & 1 & 0   & 0   & 0   &  0    &  1    & 0 \\
                  & $p_{i_{\bar{b}}}$  & 1   & 0   & 1 & 0   & 0   & 0   &  0    &  1    & 0 \\
    $j_b$         & $w_{j_b}$          & 0   & 0   & 2 & 0   & 0   & 0   &  0    &  0    & 0 \\
                  & $p_{j_b}$          & 0   & 0   & 2 & 0   & 0   & 0   &  0    &  0    & 0 \\
    $j_b^\prime$  & $w_{j_b^\prime}$   & 0   & 0   & 1 & 0   & 0   & 0   &  0    &  0    & 0 \\
                  & $p_{j_b^\prime}$   & 0   & 0   & 0 & 1   & 0   & 0   &  0    &  0    & 0 \\
    $i_c$         & $w_{i_c}$          & 0   & 0   & 0 & 0   & 1   & 0   &  0    &  0    & 1 \\ 
                  & $p_{i_c}$          & 0   & 0   & 0 & 0   & 1   & 0   &  0    &  0    & 1 \\ 
    $i_{\bar{c}}$ & $w_{i_{\bar{c}}}$  & 0   & 0   & 0 & 0   & 1   & 0   &  1    &  0    & 1 \\
                  & $p_{i_{\bar{c}}}$  & 0   & 0   & 0 & 0   & 1   & 0   &  1    &  0    & 1 \\
    $i_d$         & $w_{i_d}$          & 0   & 0   & 0 & 0   & 0   & 1   &  0    &  1    & 0 \\
                  & $p_{i_d}$          & 0   & 0   & 0 & 0   & 0   & 1   &  0    &  1    & 0 \\
    $i_{\bar{d}}$ & $w_{i_{\bar{d}}}$  & 0   & 0   & 0 & 0   & 0   & 1   &  0    &  0    & 0 \\
                  & $p_{i_{\bar{d}}}$  & 0   & 0   & 0 & 0   & 0   & 1   &  0    &  0    & 0 \\
    $i^1_{c_1}$   & $w_{i^1_{c_1}}$    & 0   & 0   & 0 & 0   & 0   & 0   & 1     & 0     & 0 \\
                  & $p_{i^1_{c_1}}$    & 0   & 0   & 0 & 0   & 0   & 0   & 1     & 0     & 0 \\
    $i^2_{c_1}$   & $w_{i^2_{c_1}}$    & 0   & 0   & 0 & 0   & 0   & 0   & 2     & 0     & 0 \\
                  & $p_{i^2_{c_1}}$    & 0   & 0   & 0 & 0   & 0   & 0   & 2     & 0     & 0 \\
    $i^1_{c_2}$   & $w_{i^1_{c_2}}$    & 0   & 0   & 0 & 0   & 0   & 0   & 0     & 1     & 0 \\
                  & $p_{i^1_{c_2}}$    & 0   & 0   & 0 & 0   & 0   & 0   & 0     & 1     & 0 \\
    $i^2_{c_2}$   & $w_{i^2_{c_2}}$    & 0   & 0   & 0 & 0   & 0   & 0   & 0     & 2     & 0 \\
                  & $p_{i^2_{c_2}}$    & 0   & 0   & 0 & 0   & 0   & 0   & 0     & 2     & 0 \\
    $i^1_{c_3}$   & $w_{i^1_{c_3}}$    & 0   & 0   & 0 & 0   & 0   & 0   & 0     & 0     & 1 \\
                  & $p_{i^1_{c_3}}$    & 0   & 0   & 0 & 0   & 0   & 0   & 0     & 0     & 1 \\
    $i^2_{c_3}$   & $w_{i^2_{c_3}}$    & 0   & 0   & 0 & 0   & 0   & 0   & 0     & 0     & 2 \\
                  & $p_{i^2_{c_3}}$    & 0   & 0   & 0 & 0   & 0   & 0   & 0     & 0     & 2 \\ \midrule
                  & $W$                & 0   & 1   & 2 & 0   & 1   & 1   & 4     & 4     & 4  \\
                  & $\bar{K}$          & 3   & 1   & 1 & 1   & 1   & 1   & 4     & 4     & 4  
    \end{tabular}
    }
    \caption{Example of construction of UTIK from an instance $\mathcal B_3 \cap 3CNF$ with $E =(a \lor b \lor \neg c) \land (\neg a \lor \neg b \lor d) \land (a \lor c \lor b) $, where $X = \{a\}$, $Y=\{ b \}$, $Z = \{c,d\}$ and the clauses are labelled from left to right.}
    \label{B3_utik}
\end{figure}

Analysing an instance from the above reduction, we can infer the following properties:
\begin{prop}\label{prop1_UTIK}
The defender can reach the profit goal $\bar K$ only fortifying either $i_x$ or $i_{\bar x}$ for each $x\in X$.
\end{prop}
\begin{proof}
First of all, if the defender wants to reach the profit goal, it is mandatory to fortify $|X|$ items in $I_X$. Otherwise, the attacker can interdict both $i_x$ and $i_{\bar x}$ for a given $x\in X$, as long as $|Y|>1$
, and forbid the defender from reaching value $(|Y|+1)|X|$ in the digits from $M$ on, and thus the profit goal.

Consider the most significant digit whose label is in $X$. If the defender fortifies neither $i_x$ nor $i_{\bar x}$, as stated above, the attacker can prevent the achievement of the profit goal $\bar K$. However, if both $i_x$ and $i_{\bar x}$ are fortified, given the fortification budget, there is a variable $x'\in X$ with $x'\neq x$ for which 
neither $i_{x'}$ nor $i_{\bar x'}$ can be fortified: consider the one with the most significant digit position. Consequently, as explained above, the attacker will interdict both such items and the defender, being unable to select more than one $I_X$ item for each position higher than $x'$,  will not reach $\bar K$. The reasoning can be propagated by induction to each variable of $X$.
\end{proof}
\begin{prop}\label{prop2_UTIK}
If the defender fortifies either $i_x$ or $i_{\bar x}$ for each $x\in X$, any solution $I_2$ for the attacker which does not select either $i_y$ or $i_{\bar y}$ for $y\in Y$ allows the defender to automatically reach the profit goal $\bar K$.
\end{prop}
\begin{proof}
As stated in the proof of Property \ref{prop1_UTIK}, once exactly one item between $i_x$ and $i_{\bar x}$ is fortified by the defender, the attacker has no incentive to attack the remaining unfortified items from $I_X$ as: i) the defender cannot select more than the fortified items from $I_X$ at the lowest decision level due to the structure of $W$; ii)  since all $|X|$ fortified items are selected at the lowest level, if the attacker does not forbid exactly $|Y|$ items from $I_Y$, the defender can select all uninterdicted $I_Y$ items and achieve automatically the profit $\bar K$. For an attacker's solution which does not interdict any item from $I_X$, the conditions for Property \ref{prop1_UBIK} apply, which completes the proof.
\end{proof}
\begin{prop}\label{prop3_UTIK}
If the defender fortifies either $i_x$ or $i_{\bar x}$ for each $x\in X$ and the attacker interdicts either $i_y$ or $i_{\bar y}$ for each $y\in Y$, 
in order to reach the profit goal $\bar K$ the follower must necessarily select in $I_1$ the $|X|$ fortified items of $I_X$, the remaining $|Y|$ uninterdicted items of $I_Y$, all $j'_{y}$ items and either $i_z$ or $i_{\bar z}$ for $z\in Z$.
\end{prop}
\begin{proof}
In order to reach the profit goal $\bar K$, the defender must select $|X|$ items from $I_X$. Moreover, due to the structure of $W$, the choice must consist of one item between $i_x$ and $i_{\bar x}$ for each $x\in X$. Indeed, one can see that for the highest $X$-related position, the defender must select only one item due to the budget limit. After having selected this item, the budget for this specific column has no residual capacity and the reasoning can be applied iteratively down to the lowest $X$-related digit. At which point, we are in a situation where Property~\ref{prop2_UBIK} can be invoked, completing the proof.
\end{proof}
\vspace{0.3cm}
\begin{thm}
UTIK is \Sth-complete.
\end{thm}
\begin{proof}
The statement of UTIK is of the form $\exists I_3 \ \ \forall I_2 \ \ \exists I_1 \ \ Q(I_1,I_2,I_3)$ where $Q$ can be tested in polynomial time, directly implying that it is in \Sth.

Let $\mathcal B_3 \cap 3CNF$ be a \emph{Yes} instance. We form a solution to UTIK by packing in $I_1$ the items $i_x$ such that $x \in X$ is 1 and the items $i_{\bar{x}}$, otherwise. Clearly, the cardinality of $I_1$ is equal to $|X|$. We can then apply Property~\ref{prop2_UTIK} to justify that either UTIK is automatically a \emph{Yes} instance or the attacker interdicts exactly one item between $i_y$ and $i_{\bar y}$ for each $y\in Y$. At which point, the defender selects exactly one item between $i_z$ and $i_{\bar z}$ in $I_1$ for each $z\in Z$ by virtue of Property~\ref{prop3_UTIK}, otherwise the instance is automatically a \emph{No} instance. 
Finally, $I_1$ is completed by items from $I_C$. We can easily map any choice of set $I_2$ to a choice of variables $Y$ in the SAT problem by setting a variable $y\in Y$ at 1 if item $i_{\bar y}$ is chosen and 0 if instead item $i_y$ is chosen. For a given set $I_2$, we can also map any set $I_1$ to a valid choice of variables from $Z$ in the SAT problem. Since the $\mathcal B_3 \cap 3CNF$ instance is a \emph{Yes} instance, it follows that the corresponding UTIK instance is also a \emph{Yes} instance, as it is possible, for any position $c\in C$, to select an item from $I_U$ which will provide a non-zero contribution in column $c$, which can then be exactly upgraded to value 4 by selecting items from $I_C$.

Next, suppose that UTIK is a \emph{Yes} instance. From Property~\ref{prop1_UTIK} we know that  $I_3$ must contain exactly one of the items $i_x$ and $i_{\bar{x}}$ for each $x \in X$. It follows, by virtue of Property~\ref{prop2_UTIK}, that the non-trivial interdiction solution $I_2$ contains exactly one item between $i_y$ and $i_{\bar y}$ for each $y\in Y$. Finally, by invoking Property~\ref{prop3_UTIK}, we know that solution $I_1$ contains exactly one item between $i_z$ and $i_{\bar z}$ for each $z\in Z$. There is, therefore, a straightforward mapping between any non trivial $I_2$ and $Y$ on the one hand and $I_1$ and $Z$ on the other. Following a similar reasoning as in the UBIK proof, we conclude that the $\mathcal B_3 \cap 3CNF$ is a \emph{Yes} instance.
\end{proof}

\subsection{Extension to any level of the polynomial hierarchy}
\label{SubSec:MIKP}

It is possible to extend the Knapsack Interdiction game to an arbitrary number of levels. Suppose that, additionally to the lowest level, which is a regular KP, we allow for $m$ levels above it which alternate between rounds of interdiction and fortification. The full problem is, therefore, an $(m+1)$-level optimisation problem, where the case with $m=1$ is the BIKP and the case with $m=2$ is the TIKP. If $m$ is odd, the highest level is an interdiction round, while if $m$ is even it is a fortification round. As in the previous subsections, we suppose that all fortification and interdiction costs are unitary. Moreover, an item which is fortified or interdicted in a certain decision round remains so in the subsequent ones and a fortified item cannot be interdicted. We formally define the following problems:
\vspace{0.3cm}
\boxxx{
\textsc{\textbf{Unitary Multi-level Interdiction Knapsack (UMIK)}}: \\
{\sc instance}: A set of items $I$ such that each $i \in I$ has a positive integer weight $w_i$ and a positive integer profit $p_i$, a positive integer capacity $W$ and a positive integer profit goal $\bar{K}$, as well as $m$ positive integer budgets $B_2,\dots,B_{m+1}$ for attack and fortification, respectively.\\
{\sc question}: $\exists I_{m+1}\forall I_m...\forall I_{1}$ (for $m$ even) or $\exists I_{m+1}\forall I_m...\exists I_{1}$ (for $m$ odd) with $|I_l|\leq B_l$ for $l\geq 2$, $\sum_{i \in I_1} w_i \leq W$, $I_m\subseteq I\setminus I_{m+1}$, $I_{m-1}\subseteq I\setminus I_m$,..., $I_1\subseteq I\setminus\cup_{l=1}^{\lfloor\frac{m+1}{2}\rfloor}I_{2l}$ such that $\sum_{i \in I_1} p_i \geq \bar{K}$ (if $m$ is even) or $\sum_{i \in I_1} p_i < \bar{K}$ (if $m$ is odd) holds?
}
\vspace{0.3cm}
In order to prove the $\Sigma_{m+1}^p$-completeness of the above problem, we will use the following problems, known to be $\Sigma_{m+1}^p$-complete \cite{Stockmeyer1976,Wrathall1976}:
\vspace{0.3cm}
\boxxx{
\textbf{\textsc{$m+1$-Alternating Quantified Satisfiability}} ($\mathcal B_{m+1} \cap \overline{3CNF}$ with $m$ odd, or $\mathcal B_{m+1} \cap 3CNF$ with $m$ even):\\
{\sc instance}: Disjoint non-empty sets of variables $X_l$ for $l=1,...,m+1$, and a Boolean expression  $E$ over $\cup_{l=1}^{m+1} X_l$ in conjunctive normal form with at most 3 literals in each clause $c \in C$. \\
{\sc question}: $\exists X_{m+1}\forall X_m...\forall X_{1}$: $E$ is never satisfied? (for $m$ odd) or $\exists X_{m+1}\forall X_m...\exists X_{1}$: $E$ is satisfied? (for $m$ even)
}

We state the following:

\begin{thm}\label{thm:UMIK}
UMIK with $m$ alternate rounds of fortification and interdiction and with unit fortification and attack costs is $\Sigma_{m+1}^p$-complete.
\end{thm}
Since it uses mainly ingredients introduced in the previous proofs, the proof is relegated to Appendix~\ref{app:UMIKP}.

We can make an interesting general remark regarding the UMIKP for any number of fortification and interdiction rounds:
\begin{prop}
When either the profits or the weights of the lowest-level KP become unitary, UMIKP becomes solvable in polynomial time.
\end{prop}
This is straightforward since any round of fortification or attack will concentrate on those items not already fortified or interdicted with either higher profit (for unit weights) or higher weights (for unit profits) so the choices at each level are straightforward. We conclude that there is therefore no intermediate situation in this regard, the problems are either complete for their natural complexity class in the polynomial hierarchy, or they become polynomial. This, of course, does not take into account the possibility of having specially structured weight or profit sets which are not simply all unitary.

\section{The Max-Flow Interdiction Problem with Fortification}
\label{sec:MaxFlow}

Given that the MFIP is the prototype of an NP-complete interdiction problem \cite{Wood1993} and many works exist on network interdiction which are based on some kind of network flow~\cite{Cormican1998,Enayaty-Ahangar2019,Kuttler2024,Sullivan2014}, we propose to study a fortification version of this problem which can serve as a basis for studying the complexity of flow-based fortification-interdiction problems. We therefore start this section by defining a straightforward extension of the MFIP to a fortification tri-level version and proceed to demonstrate that this problem is complete for the second level of the polynomial hierarchy, even considering unit fortification and attack costs.

\paragraph{The Max-Flow Interdiction Problem with Fortification}
The MFIPF is defined over a directed graph $G=(V,A)$, where each arc $(i,j)\in A$ is associated with a maximum capacity $c_{ij}$, an attack cost $w_{ij}$ and a fortification cost $b_{ij}$. The aim of the defender is to maximise the amount of flow through the graph, from a source node $s$ to a destination node $t$. The attacker selects a subset of arcs $D$ to destroy, given an attack budget $W$ (that is, $\sum_{(i, j) \in D} w_{i j} \leq W$), in order to minimise the maximum flow from $s$ to $t$ on the graph induced by the arc removal. Before the attack, the defender can take preventive actions that consist, given a fortification budget $B$, in selecting a subset of arcs $F$ which become immune to any attack (that is, $\sum_{(i, j) \in F} b_{i j} \leq B$). The unitary version of the problem considers only unit attack and fortification costs.

\paragraph{\Stwo-completeness of the MFIPF with unit costs}
We now proceed to demonstrate the completeness of the MFIPF with unit costs in the second level of the polynomial hierarchy.
\vspace{0.3cm}
\boxxx{
\textsc{\textbf{Unitary Max-Flow Interdiction Problem with Fortification (UMFIPF)}}: \\
{\sc instance}: A directed graph $G=(V,A)$ with a maximum integer capacity $c_{ij}$ for each $(i,j)\in A$, a source and a destination node $s,t\in V$, a positive integer flow goal $\bar{K}$ and two positive integer budgets $W$ and $B$ for attack and fortification.\\
{\sc question}: Is there a subset $F\subseteq A$ of arcs, with $|F| \leq B$, such that for every subset $D\subseteq A\setminus F$, with $|D| \leq W$, the value of the maximum flow from $s$ to $t$ in the induced graph $G_D=(V,A\setminus D)$ is at least $\bar{K}$?
}
\vspace{0.3cm}
In order to prove the \Stwo-completeness of the above problem, we will use {\sc Clique Interdiction}, known to be \Stwo-complete~\cite{Rutenburg1994}:
\vspace{0.3cm}
\boxxx{
\textbf{\textsc{Clique Interdiction}}:\\
{\sc instance}: Simple graph $H=(N,E)$ and two positive integers $\ell \geq 4$ and $r$, where $|E| \geq \binom{\ell}{2} $. \\
{\sc question}: Is there $R \subseteq N$ with $|R| \leq r$ such that for every clique $Q$ of $H$ with $|Q|=\ell$, we have $R\cap Q \neq \emptyset$.
}
\vspace{0.3cm}
Note that the two constraints in the input are without loss of generality. If $|E| < \binom{\ell}{2}$, we can add isolated edges to $H$ (hence increasing the cardinality of $|E|$) which does not contribute to a clique of size $\ell\geq 4$. For the condition $\ell \geq 4$, note that we can always add four graphs, $G_1=(\{v_{1,1},\ldots,v_{1,r},v_{1,r+1}\}, \emptyset), \ldots,$ and $ G_4=(\{v_{4,1},\ldots,v_{4,r},v_{4,r+1}\}, \emptyset),$ to $H$ and add edges between them and all vertices of $H$, and between all vertices of $G_i$ and $G_j$ for $i \neq j$. In this way, the maximum clique of the modified graph is exactly the maximum clique of $H$ plus 4. Moreover, if we interdict a set of vertices $R$ of the modified graph, its maximum clique is the same as that of $H(N\setminus (R\cap N))$\footnote{Subgraph of $H$ induced by nodes $N\setminus (R\cap N)$.} plus 4.

Now we build our reduction. From an instance $(H=(N,E), \ell, r)$ of {\sc Clique interdiction}, we build an instance $(G=(V,A), s,t,c, \bar K, B, W)$ of UMFIPF :
\begin{itemize}
    \item The set of vertices is $V=\{s,t\} \cup \{z_e: e \in E\} \cup \{v_i: i \in N\}$.
    \item The set of arcs $A$ consists of $A_1=\{(s,z_e): e \in E\}$ with unitary capacities, $A_2=\{ (z_e, v_i): e\in E \wedge i \in e \}$ with unitary capacities and $A_3=\{(v_i,t): i \in N \}$ with capacity $|E|=m$.
    \item $B=r$, $W=\ell$ and $\bar K = m-  \binom{\ell}{2} +1$ (note that $\bar K \geq 1$ since $m \geq \binom{\ell}{2}$).
\end{itemize}
See Figure~\ref{fig:construction} for an illustration of the reduction.
\tikzset{
  hvtx/.style   = {circle, draw, thin, minimum size=5.5mm, inner sep=0pt,
                   font=\footnotesize},
  hvtxU/.style  = {hvtx, fill=black!12},
  hedge/.style  = {draw=black!55, thin},
  hedgeU/.style = {draw=black, line width=1.1pt},
  znode/.style  = {rectangle, draw, thin, rounded corners=1.5pt,
                   minimum width=9mm, minimum height=5.2mm, inner sep=1pt,
                   font=\scriptsize},
  zdead/.style  = {znode, fill=black!10, draw=black!35, text=black!45},
  zlive/.style  = {znode, line width=1.1pt},
  vnode/.style  = {circle, draw, thin, minimum size=5.5mm, inner sep=0pt,
                   font=\footnotesize},
  vnodeU/.style = {vnode, fill=black!12},
  term/.style   = {circle, draw, thin, minimum size=6mm, inner sep=0pt,
                   font=\footnotesize},
  arc/.style    = {-{Stealth[length=1.5mm]}, draw=black!45, thin},
  arcdead/.style= {-{Stealth[length=1.5mm]}, draw=black!20, thin},
  arclive/.style= {-{Stealth[length=1.5mm]}, draw=black, line width=1pt},
  del/.style    = {draw=black, dashed, thin},
  xmark/.style  = {inner sep=0.5pt, fill=white, font=\scriptsize},
}
\begin{figure}[htbp]
\centering
\begin{subfigure}[b]{0.30\textwidth}
\centering
\begin{tikzpicture}[scale=1.0]
  \node[hvtx] (h1) at (0,1.4)    {$1$};
  \node[hvtx] (h2) at (1.4,1.4)  {$2$};
  \node[hvtx] (h3) at (0,0)      {$3$};
  \node[hvtx] (h4) at (1.4,0)    {$4$};
  \node[hvtx] (h5) at (2.7,-0.8) {$5$};
  \draw[hedge] (h1) -- (h2);  \draw[hedge] (h1) -- (h3);
  \draw[hedge] (h1) -- (h4);  \draw[hedge] (h2) -- (h3);
  \draw[hedge] (h2) -- (h4);  \draw[hedge] (h3) -- (h4);
  \draw[hedge] (h4) -- (h5);
\end{tikzpicture}
\caption{The graph $H$, with $|E|=m=7$.}
\label{fig:constr-H}
\end{subfigure}
\hfill
\begin{subfigure}[b]{0.66\textwidth}
\centering
\begin{tikzpicture}[scale=0.72]
  \node[term]  (s)   at (-2.6,0) {$s$};
  \node[znode] (z12) at (0, 3)   {$z_{12}$};
  \node[znode] (z13) at (0, 2)   {$z_{13}$};
  \node[znode] (z14) at (0, 1)   {$z_{14}$};
  \node[znode] (z23) at (0, 0)   {$z_{23}$};
  \node[znode] (z24) at (0,-1)   {$z_{24}$};
  \node[znode] (z34) at (0,-2)   {$z_{34}$};
  \node[znode] (z45) at (0,-3)   {$z_{45}$};
  \node[vnode] (v1)  at (3.4, 2) {$v_1$};
  \node[vnode] (v2)  at (3.4, 1) {$v_2$};
  \node[vnode] (v3)  at (3.4, 0) {$v_3$};
  \node[vnode] (v4)  at (3.4,-1) {$v_4$};
  \node[vnode] (v5)  at (3.4,-2) {$v_5$};
  \node[term]  (t)   at (6.0,0)  {$t$};
  \foreach \z in {z12,z13,z14,z23,z24,z34,z45} \draw[arc] (s) -- (\z);
  \draw[arc] (z12) -- (v1);  \draw[arc] (z12) -- (v2);
  \draw[arc] (z13) -- (v1);  \draw[arc] (z13) -- (v3);
  \draw[arc] (z14) -- (v1);  \draw[arc] (z14) -- (v4);
  \draw[arc] (z23) -- (v2);  \draw[arc] (z23) -- (v3);
  \draw[arc] (z24) -- (v2);  \draw[arc] (z24) -- (v4);
  \draw[arc] (z34) -- (v3);  \draw[arc] (z34) -- (v4);
  \draw[arc] (z45) -- (v4);  \draw[arc] (z45) -- (v5);
  \foreach \v in {v1,v2,v3,v4,v5} \draw[arc] (\v) -- (t);
  \node[font=\scriptsize, align=center] at (-1.6,3.4) {$A_1$\\[-1pt]cap.\ $1$};
  \node[font=\scriptsize, align=center] at ( 1.7,3.4) {$A_2$\\[-1pt]cap.\ $1$};
  \node[font=\scriptsize, align=center] at ( 5.0,3.4) {$A_3$\\[-1pt]cap.\ $m$};
\end{tikzpicture}
\caption{The network $G$ built from $H$.}
\label{fig:constr-G}
\end{subfigure}
\caption{The reduction, on the instance $H$ with $\ell=4$,
so that $m=7\geq\binom{\ell}{2}=6$ and $\bar K=2$. Every edge $e\in E$
becomes an edge node $z_e$ whose two outgoing arcs lead to the two
endpoints of $e$; every vertex $i\in N$ becomes a vertex node $v_i$ whose
\emph{only} outgoing arc is $(v_i,t)$.}
\label{fig:construction}
\end{figure}

To complete the proof, we make use of several lemmata, allowing us to lower and upper-bound the maximum flow of a graph. Before stating them, we introduce some definitions. For  $D \subseteq A$, we define $U(D)=\{ i \in N: (v_i,t) \in D\}$. For $U \subseteq N$, we define $E(U)$ as the set of edges in $H$ with both end points in $U$. An edge $e=\{i,j\} \in E$ is called \emph{alive under $D$} if $(s,z_e) \notin D$, there is $i \in e$ such that $(z_e,v_i) \notin D$ and $i \notin U(D)$ (i.e., $(v_i,t)\notin D$); otherwise, it is designated as \emph{dead under $D$}. See Figure~\ref{fig:definitions} for an illustration of these definitions.
\begin{figure}[htbp]
\centering

\begin{subfigure}[b]{\textwidth}
\centering
\begin{tikzpicture}[scale=0.78]
  \node[term]  (s)   at (-2.6,0) {$s$};
  \node[znode] (z12) at (0, 3)   {$z_{12}$};
  \node[znode] (z13) at (0, 2)   {$z_{13}$};
  \node[znode] (z14) at (0, 1)   {$z_{14}$};
  \node[znode] (z23) at (0, 0)   {$z_{23}$};
  \node[znode] (z24) at (0,-1)   {$z_{24}$};
  \node[znode] (z34) at (0,-2)   {$z_{34}$};
  \node[znode] (z45) at (0,-3)   {$z_{45}$};
  \node[vnodeU](v1)  at (3.4, 2) {$v_1$};
  \node[vnodeU](v2)  at (3.4, 1) {$v_2$};
  \node[vnode] (v3)  at (3.4, 0) {$v_3$};
  \node[vnode] (v4)  at (3.4,-1) {$v_4$};
  \node[vnode] (v5)  at (3.4,-2) {$v_5$};
  \node[term]  (t)   at (6.0,0)  {$t$};
  \foreach \z in {z12,z13,z14,z23,z24,z45} \draw[arc] (s) -- (\z);
  \draw[del] (s) -- node[xmark,pos=0.62]{$\times$} (z34);
  \draw[arc] (z12) -- (v1);  \draw[arc] (z12) -- (v2);
  \draw[arc] (z13) -- (v1);
  \draw[del] (z13) -- node[xmark,pos=0.55]{$\times$} (v3);
  \draw[arc] (z14) -- (v1);  \draw[arc] (z14) -- (v4);
  \draw[arc] (z23) -- (v2);  \draw[arc] (z23) -- (v3);
  \draw[arc] (z24) -- (v2);  \draw[arc] (z24) -- (v4);
  \draw[arc] (z34) -- (v3);  \draw[arc] (z34) -- (v4);
  \draw[arc] (z45) -- (v4);  \draw[arc] (z45) -- (v5);
  \draw[del] (v1) -- node[xmark,pos=0.55]{$\times$} (t);
  \draw[del] (v2) -- node[xmark,pos=0.55]{$\times$} (t);
  \foreach \v in {v3,v4,v5} \draw[arc] (\v) -- (t);
  \begin{scope}[on background layer]
    \node[draw, dashed, rounded corners, inner sep=3pt, fit=(v1)(v2)] {};
  \end{scope}
  \node[font=\footnotesize] at (3.4,3.05) {$U(D)=\{1,2\}$};
\end{tikzpicture}
\caption{The attack $D=\{(v_1,t),(v_2,t),(s,z_{34}),(z_{13},v_3)\}$, with
$|D|=4=W$. Dashed arcs marked $\times$ are the arcs of $D$; the set $U(D)$
collects the vertices whose arc to $t$ lies in $D$.}
\label{fig:def-UD}
\end{subfigure}

\vspace{0.6cm}

\begin{subfigure}[b]{0.36\textwidth}
\centering
\begin{tikzpicture}[scale=1.0]
  \node[hvtxU] (h1) at (0,1.4)    {$1$};
  \node[hvtxU] (h2) at (1.4,1.4)  {$2$};
  \node[hvtx]  (h3) at (0,0)      {$3$};
  \node[hvtx]  (h4) at (1.4,0)    {$4$};
  \node[hvtx]  (h5) at (2.7,-0.8) {$5$};
  \draw[hedgeU] (h1) -- (h2);
  \draw[hedge]  (h1) -- (h3);  \draw[hedge] (h1) -- (h4);
  \draw[hedge]  (h2) -- (h3);  \draw[hedge] (h2) -- (h4);
  \draw[hedge]  (h3) -- (h4);  \draw[hedge] (h4) -- (h5);
  \begin{scope}[on background layer]
    \node[draw, dashed, rounded corners, inner sep=4pt, fit=(h1)(h2)] {};
  \end{scope}
  \node[font=\footnotesize] at (0.7,2.35) {$U(D)$};
\end{tikzpicture}
\caption{$E(U(D))=\{\,\{1,2\}\,\}$: the edges of $H$ with \emph{both}
endpoints in $U(D)$.}
\label{fig:def-EU}
\end{subfigure}
\hfill
\begin{subfigure}[b]{0.60\textwidth}
\centering
\begin{tikzpicture}[scale=0.66]
  \node[term]  (s)   at (-2.6,0) {$s$};
  \node[zdead] (z12) at (0, 3)   {$z_{12}$};
  \node[zdead] (z13) at (0, 2)   {$z_{13}$};
  \node[zlive] (z14) at (0, 1)   {$z_{14}$};
  \node[zlive] (z23) at (0, 0)   {$z_{23}$};
  \node[zlive] (z24) at (0,-1)   {$z_{24}$};
  \node[zdead] (z34) at (0,-2)   {$z_{34}$};
  \node[zlive] (z45) at (0,-3)   {$z_{45}$};
  \node[vnodeU](v1)  at (3.4, 2) {$v_1$};
  \node[vnodeU](v2)  at (3.4, 1) {$v_2$};
  \node[vnode] (v3)  at (3.4, 0) {$v_3$};
  \node[vnode] (v4)  at (3.4,-1) {$v_4$};
  \node[vnode] (v5)  at (3.4,-2) {$v_5$};
  \node[term]  (t)   at (6.0,0)  {$t$};
  \draw[arcdead] (s) -- (z12);
  \draw[arcdead] (s) -- (z13);
  \draw[arclive] (s) -- (z14);
  \draw[arclive] (s) -- (z23);
  \draw[arclive] (s) -- (z24);
  \draw[del]     (s) -- node[xmark,pos=0.62]{$\times$} (z34);
  \draw[arclive] (s) -- (z45);
  \draw[arcdead] (z12) -- (v1);  \draw[arcdead] (z12) -- (v2);
  \draw[arcdead] (z13) -- (v1);
  \draw[del]     (z13) -- node[xmark,pos=0.55]{$\times$} (v3);
  \draw[arcdead] (z14) -- (v1);  \draw[arclive] (z14) -- (v4);
  \draw[arcdead] (z23) -- (v2);  \draw[arclive] (z23) -- (v3);
  \draw[arcdead] (z24) -- (v2);  \draw[arclive] (z24) -- (v4);
  \draw[arcdead] (z34) -- (v3);  \draw[arcdead] (z34) -- (v4);
  \draw[arclive] (z45) -- (v4);  \draw[arcdead] (z45) -- (v5);
  \draw[del] (v1) -- node[xmark,pos=0.55]{$\times$} (t);
  \draw[del] (v2) -- node[xmark,pos=0.55]{$\times$} (t);
  \draw[arclive] (v3) -- (t);
  \draw[arclive] (v4) -- (t);
  \draw[arcdead] (v5) -- (t);
\end{tikzpicture}
\caption{Alive (bold) and dead (grey) edge vertices under $D$. Bold arcs carry
the flow of Lemma~\ref{lemma1:fix}, of value $4$.}
\label{fig:def-alive}
\end{subfigure}

\caption{The definitions $U(D)$, $E(U(D))$ and alive/dead on the running
example. Edge $\{1,2\}$ is dead because both endpoints lie in $U(D)$; edge
$\{3,4\}$ is dead because its own arc $(s,z_{34})$ was deleted; edge
$\{1,3\}$ is dead because one endpoint lies in $U(D)$ and the arc to the
other endpoint was deleted. The four remaining edges stay alive through
$v_3$, $v_4$ or $v_5$. The three dead edges meet the bound of Lemma~\ref{lemma2:fix} with
equality: $|E(U(D))|+(\ell-|U(D)|)=1+2=3$.}
\label{fig:definitions}
\end{figure}

\begin{lem}\label{lemma1:fix}
For all $D \subseteq A$, let $G_D$ denote graph $G$ after interdicting (deleting) arcs in $D$. The maximum flow on the graph $G_D$  is greater or equal to the number of alive edges under $D$. 
$$maxflow(G_D) \geq |\{\textrm{edges alive under }D \}|$$
\end{lem}
\begin{proof}{Proof}
    For each alive edge $e=\{i,j\}$ under $D$, the following arcs are not interdicted: arc $(s,z_e)$, arc $(z_e,v_i)$ (or $(z_e,v_j)$) and arc $(v_i,t)$ (or $(v_j,t)$). Hence, we can send one unit of flow through the path $(s,z_e,v_i,t)$ (or $(s,z_e,v_j,t)$). Since we can do this for each of the alive edges under $D$, the stated result follows.
\end{proof}

\begin{lem}\label{lemma2:fix}
For every $D \subseteq A$ with $|D|\leq \ell$, the number of dead edges
under $D$ is at most $|E(U(D))|+(\ell-|U(D)|)$.
$$|E(U(D))|+(\ell-|U(D)|) \geq |\{\textrm{dead edges under }D\}|$$
\end{lem}
\begin{proof}{Proof}
Let $e=\{i,j\} \in E$ be dead under $D$. If $i,j \in U(D)$, then
$e \in E(U(D))$. Otherwise, one endpoint, say $i$, lies outside $U(D)$;
if we had both $(s,z_e) \notin D$ and $(z_e,v_i) \notin D$, then $i$ would
make $e$ alive, so $D$ contains $(s,z_e)$ or $(z_e,v_i)$.
Both arcs are incident to $z_e$, and the vertices $z_e$ are distinct for
distinct edges, so distinct dead edges outside $E(U(D))$ are mapped to
distinct arcs of $D \setminus A_3$. Since $|D \cap A_3| = |U(D)|$, there
are at most $\ell - |U(D)|$ such arcs, and adding the at most
$|E(U(D))|$ dead edges inside $E(U(D))$ gives the bound.
\end{proof}

The next lemmata allow us to connect a clique with an attack in our maximum flow problem; see Figure~\ref{fig:clique-attack}.

\begin{lem}\label{lemma3:fix}
Let $\ell \geq 4$ and let $U \subseteq N$ with $u=|U| \leq \ell$. Then
\[
  |E(U)| + (\ell - u) \;\leq\; \binom{\ell}{2},
\]
with equality if and only if $u=\ell$ and $U$ is a clique of $H$.
In particular, if $u<\ell$ or $U$ is not a clique of $H$, then
$|E(U)|+(\ell-u) \leq \binom{\ell}{2}-1$.
\end{lem}
\begin{proof}{Proof}
We have $|E(U)| \leq \binom{u}{2}$, with equality if and only if $U$ is a
clique of $H$. Hence, it suffices to bound $g(u) := \binom{u}{2} + \ell - u$.
Since $\binom{\ell}{2}-\binom{u}{2} = \frac{(\ell-u)(\ell+u-1)}{2}$,
\[
  \binom{\ell}{2} - g(u)
  \;=\; \frac{(\ell-u)(\ell+u-1)}{2} - (\ell-u)
  \;=\; \frac{(\ell-u)(\ell+u-3)}{2}.
\]
If $u=\ell$ this vanishes, and if $u<\ell$ both factors are positive because
$\ell-u \geq 1$ and $\ell+u-3 \geq \ell-3 \geq 1$. Hence $g(u) \leq
\binom{\ell}{2}$ for every $u \leq \ell$, with equality only for $u=\ell$,
which proves the bound.

For the equality case, $|E(U)|+(\ell-u) \leq g(u) \leq \binom{\ell}{2}$, so
equality throughout forces $g(u)=\binom{\ell}{2}$, hence $u=\ell$, and
$|E(U)|=\binom{u}{2}$, implying that $U$ is a clique. The converse is immediate. The
last assertion follows since all quantities are integers. 
\end{proof}

\begin{lem}\label{lemma4:fix}
Let $Q \subseteq N$ be a clique of $H$ with $|Q|=\ell$, and let
$D_Q = \{(v_i,t) : i \in Q\}$, so that $|D_Q|=\ell=W$ and $U(D_Q)=Q$. Then
\[
  maxflow(G_{D_Q}) \;\leq\; m - \binom{\ell}{2} \;<\; \bar K .
\]
\end{lem}
\begin{proof}{Proof}
Intuitively, deleting $D_Q$ strands every edge vertex $z_e$ with $e\in E(Q)$,
because the only two arcs leaving $z_e$ lead to $v_i$ and $v_j$, whose only
arcs to $t$ have just been deleted; the flow is then limited by the arcs of
$A_1$ belonging to the remaining edges. We turn this into an $s$--$t$ cut.

Let $e=\{i,j\} \in E(Q)$. The arcs leaving $z_e$ are exactly $(z_e,v_i)$ and
$(z_e,v_j)$, and the only arc leaving $v_k$, for any $k \in N$, is $(v_k,t)$.
Since $i,j \in Q$, both $(v_i,t)$ and $(v_j,t)$ belong to $D_Q$, so no path of
$G_{D_Q}$ leaves $z_e$ and reaches $t$.

Consider now $C = \{(s,z_e) : e \in E \setminus E(Q)\}$ and let $P$ be any
$s$--$t$ path of $G_{D_Q}$. The first arc of $P$ leaves $s$, hence is of the
form $(s,z_e)$ for some $e \in E$, as $A_1$ contains every arc leaving $s$.
By the previous paragraph $e \notin E(Q)$, and therefore $P$ uses an arc of
$C$. Consequently $C$ is an $s$--$t$ cut of $G_{D_Q}$.

Every arc of $A_1$ has unit capacity, so
\[
  cap(C) \;=\; |E \setminus E(Q)| \;=\; m - |E(Q)| \;=\; m - \binom{\ell}{2},
\]
with the last equality comes from $Q$ being a clique of      $\ell$ vertices. By the max-flow min-cut theorem, the value of any
flow is bounded above by the capacity of any $s$--$t$ cut, whence
$maxflow(G_{D_Q}) \leq m - \binom{\ell}{2} < m - \binom{\ell}{2} + 1 = \bar K$.
\end{proof}
\begin{figure}[htbp]
\centering
\begin{subfigure}[b]{0.36\textwidth}
\centering
\begin{tikzpicture}[scale=1.0]
  \node[hvtxU] (h1) at (0,1.4)    {$1$};
  \node[hvtxU] (h2) at (1.4,1.4)  {$2$};
  \node[hvtxU] (h3) at (0,0)      {$3$};
  \node[hvtxU] (h4) at (1.4,0)    {$4$};
  \node[hvtx]  (h5) at (2.7,-0.8) {$5$};
  \draw[hedgeU] (h1) -- (h2);  \draw[hedgeU] (h1) -- (h3);
  \draw[hedgeU] (h1) -- (h4);  \draw[hedgeU] (h2) -- (h3);
  \draw[hedgeU] (h2) -- (h4);  \draw[hedgeU] (h3) -- (h4);
  \draw[hedge]  (h4) -- (h5);
  \begin{scope}[on background layer]
    \node[draw, dashed, rounded corners, inner sep=4pt,
          fit=(h1)(h2)(h3)(h4)] {};
  \end{scope}
  \node[font=\footnotesize] at (0.7,2.35) {$Q$, a clique of $H$};
\end{tikzpicture}
\caption{The clique $Q=\{1,2,3,4\}$, with $|Q|=\ell=4$ and
$|E(Q)|=\binom{4}{2}=6$.}
\end{subfigure}
\hfill
\begin{subfigure}[b]{0.60\textwidth}
\centering
\begin{tikzpicture}[scale=0.66]
  \node[term]  (s)   at (-2.6,0) {$s$};
  \node[zdead] (z12) at (0, 3)   {$z_{12}$};
  \node[zdead] (z13) at (0, 2)   {$z_{13}$};
  \node[zdead] (z14) at (0, 1)   {$z_{14}$};
  \node[zdead] (z23) at (0, 0)   {$z_{23}$};
  \node[zdead] (z24) at (0,-1)   {$z_{24}$};
  \node[zdead] (z34) at (0,-2)   {$z_{34}$};
  \node[zlive] (z45) at (0,-3)   {$z_{45}$};
  \node[vnodeU](v1)  at (3.4, 2) {$v_1$};
  \node[vnodeU](v2)  at (3.4, 1) {$v_2$};
  \node[vnodeU](v3)  at (3.4, 0) {$v_3$};
  \node[vnodeU](v4)  at (3.4,-1) {$v_4$};
  \node[vnode] (v5)  at (3.4,-2) {$v_5$};
  \node[term]  (t)   at (6.0,0)  {$t$};
  \foreach \z in {z12,z13,z14,z23,z24,z34} \draw[arcdead] (s) -- (\z);
  \draw[arclive] (s) -- (z45);
  \draw[arcdead] (z12) -- (v1);  \draw[arcdead] (z12) -- (v2);
  \draw[arcdead] (z13) -- (v1);  \draw[arcdead] (z13) -- (v3);
  \draw[arcdead] (z14) -- (v1);  \draw[arcdead] (z14) -- (v4);
  \draw[arcdead] (z23) -- (v2);  \draw[arcdead] (z23) -- (v3);
  \draw[arcdead] (z24) -- (v2);  \draw[arcdead] (z24) -- (v4);
  \draw[arcdead] (z34) -- (v3);  \draw[arcdead] (z34) -- (v4);
  \draw[arcdead] (z45) -- (v4);  \draw[arclive] (z45) -- (v5);
  \draw[del] (v1) -- node[xmark,pos=0.55]{$\times$} (t);
  \draw[del] (v2) -- node[xmark,pos=0.55]{$\times$} (t);
  \draw[del] (v3) -- node[xmark,pos=0.55]{$\times$} (t);
  \draw[del] (v4) -- node[xmark,pos=0.55]{$\times$} (t);
  \draw[arclive] (v5) -- (t);
\end{tikzpicture}
\caption{Only $z_{45}$ survives, so the maximum flow is $1<\bar K=2$.}
\end{subfigure}
\caption{The clique attack $D_Q=\{(v_i,t): i\in Q\}$ associated with the
clique $Q$, with $|D_Q|=|Q|=\ell=4=W$ and $U(D_Q)=Q$. Four deletions kill
six edges, which is the equality case of Lemma~\ref{lemma3:fix}; the cut of Lemma~\ref{lemma4:fix}
consists of the single arc $(s,z_{45})$. No fortification is present here,
so the budget $r$ plays no role.}
\label{fig:clique-attack}
\end{figure}

\begin{thm}\label{thm:umfipf-fix}
UMFIPF is \Stwo-complete.
\end{thm}
\begin{proof}{Proof}
Membership in \Stwo{} is immediate: the problem asks whether there exists
$F \subseteq A$ with $|F| \leq B$ such that for all $D \subseteq A\setminus F$
with $|D| \leq W$ we have $maxflow(G_D) \geq \bar K$, where $F$ and $D$ are of
polynomial size and, the maximum flow problem being solvable in polynomial
time, the innermost predicate is decidable in polynomial time.

For the hardness, we show that an instance $(H=(N,E),\ell,r)$ of
{\sc Clique Interdiction} is a YES-instance if and only if the instance
$(G=(V,A),s,t,c,\bar K,B,W)$ of UMFIPF built above is a YES-instance. The
construction takes polynomial time, since $|V|=n+m+2$ and $|A|=3m+n$, and it
produces a valid instance because $\bar K = m-\binom{\ell}{2}+1 \geq 1$ by the
restriction $m \geq \binom{\ell}{2}$ on the input. Recall also that every edge
of $H$ is either alive or dead under $D$, so the numbers of alive and of dead
edges sum to $m$.

\medskip
\noindent\emph{($\Rightarrow$)} Let $R \subseteq N$ with $|R| \leq r$ be such
that $R \cap Q \neq \emptyset$ for every clique $Q$ of $H$ with $|Q|=\ell$.
Take
\[
  F \;=\; \{(v_i,t) : i \in R\} \;\subseteq\; A_3 ,
\]
so that $|F| = |R| \leq r = B$. Let $D \subseteq A \setminus F$ with
$|D| \leq W = \ell$, and define $U = U(D)$ and $u = |U|$.

Since $(v_i,t) \in F$ for every $i \in R$ and $D \cap F = \emptyset$, we have
$U \cap R = \emptyset$. We claim that $u < \ell$ or $U$ is not a clique of $H$.
Indeed, if $u = \ell$ and $U$ were a clique of $H$, then $U$ would be a clique
of size $\ell$ with $R \cap U = \emptyset$, contradicting the choice of $R$.
Lemma~\ref{lemma3:fix} therefore applies in its strict form and gives
\[
  |E(U)| + (\ell - u) \;\leq\; \binom{\ell}{2} - 1 .
\]
By Lemma~\ref{lemma2:fix} the number of dead edges under $D$ is at most
$\binom{\ell}{2}-1$, hence the number of alive edges under $D$ is at least
$m - \binom{\ell}{2} + 1 = \bar K$, and Lemma~\ref{lemma1:fix} yields
$maxflow(G_D) \geq \bar K$. As $D$ was arbitrary, $F$ certifies that the
UMFIPF instance is a YES-instance.

\medskip
\noindent\emph{($\Leftarrow$)} Let $F \subseteq A$ with $|F| \leq B = r$ be
such that $maxflow(G_D) \geq \bar K$ for every $D \subseteq A \setminus F$ with
$|D| \leq W$. Take
\[
  R \;=\; \{ i \in N : (v_i,t) \in F \} ,
\]
so that $|R| \leq |F| \leq r$. Suppose, for the sake of contradiction, that
some clique $Q$ of $H$ with $|Q| = \ell$ satisfies $R \cap Q = \emptyset$, and
consider $D_Q = \{(v_i,t) : i \in Q\}$. For $i \in Q$ we have $i \notin R$,
hence $(v_i,t) \notin F$ by the definition of $R$; therefore
$D_Q \subseteq A \setminus F$, and $|D_Q| = \ell = W$. Lemma~\ref{lemma4:fix}
then gives $maxflow(G_{D_Q}) < \bar K$, contradicting the choice of $F$.
Consequently every clique of $H$ of size $\ell$ meets $R$, and
$(H,\ell,r)$ is a YES-instance of {\sc Clique Interdiction}.
\end{proof}

\section{The Shortest Path Interdiction Problem with Fortification}
\label{sec:ShortestPath}

The SPIP is a classic network interdiction problem which has been tackled under many different forms, ranging from the most basic one~\cite{Israeli2002} to more refined versions with uncertain and asymmetric information~\cite{Nguyen2024}. It is formulated on a directed graph with a given source node and destination node, and consists in an attacker selecting a subset of arcs subject to a certain attack budget, in order to maximise the length of the shortest path that a defender can find from the source to the destination. The fortification version of this problem, where the defender can take preventive actions by fortifying a subset of arcs subject to a fortification budget, is considered as a basic problem in fortification-interdiction games and has been used as an example problem for applying generic tri-level fortification algorithmic frameworks in several works~\cite{LEITNER20231026,Lozano2017}. The version advocated in such works is usually one with unit fortification and attack costs. In this section, we will first present the general SPIP with Fortification (SPIPF), then we will prove the \Stwo-completeness of a version of SPIPF with unit fortification costs.

\paragraph{Problem definition}
The SPIPF is defined over a directed graph $G=(V,A)$, where each arc $(i,j)\in A$ is associated with a length $d_{ij}$, an attack cost $w_{ij}$ and a fortification cost $b_{ij}$. The aim of the defender is to find a path of minimum length between a source node $s$ and a destination node $t$. The attacker seeks to interdict a set of arcs $D$, given a maximum attack budget $W'$ (i.e., $\sum_{(i, j) \in D} w_{i j} \leq W'$), in order to maximise the length of the shortest path on the graph induced by the arc removal. Before the attacker operates, the defender can take preventive actions to fortify a subset of arcs $F$, which become immune to any attack, given a maximum fortification budget $B'$ (i.e., $\sum_{(i, j) \in F} b_{i j} \leq B'$). The unitary version of that problem considers only unit fortification costs, but arbitrary attack costs.

\boxxx{
\textsc{\textbf{Shortest Path Interdiction Problem with Unitary Fortification (SPIPUF)}}: \\
{\sc instance}: A directed graph $G=(V,A)$ with an integer length $d_{ij}$ and an attack cost $w_{ij}$ for each $(i,j)\in A$, a source and destination node $s,t\in V$, two positive integer budgets $W'$ and $B'$ and a positive integer goal $\bar{K'}$.\\
{\sc question}: Is there a subset $F\subseteq A$ of arcs, with $|F| \leq B'$, such that for every subset $D\subseteq A\setminus F$, with $\sum_{(i,j)\in D}w_{ij} \leq W'$, the length of the shortest path from $s$ to $t$ in the induced graph $G_D=(V,A\setminus D)$ is strictly less than $\bar{K'}$?
}

\paragraph{\Stwo-completeness of SPIPF with unit fortification costs}
We now proceed to demonstrate the \Stwo-completeness of the SPIPF with unit fortification costs. To this purpose, we will use UBIK, demonstrated to be in the second level of the polynomial hierarchy in Section~\ref{sec:Knapsack}.

\begin{thm}
SPIPUF is \Stwo-complete.
\end{thm}
\begin{proof}
The statement of SPIPUF is of the form $\exists F \ \ \forall D \ \ Q(F,D)$, where $Q$ can be tested in polynomial time, since the Shortest Path Problem is in P, directly implying that SPIPUF is in \Stwo.

Let us introduce the following reduction from UBIK:
\begin{itemize}
\item  We label each of the items in UBIK with a number from 1 to $n=|I|$.
\item We introduce in set $V$ a node $v_k$ for each item $k\in\{1,...,n\}$ and a source node $v_0$; node $v_n$ will serve as destination node.
\item  For each item $i\in I=\{1,...,n\}$, we introduce the following gadget: add an arc $(v_{i-1},v_i)$ with length $p_i+2$ and attack cost $W+1$, a node $v_i^{\prime}$ and arcs $(v_{i-1},v_i^{\prime})$ with length 1 and attack cost $w_i$ and $(v_i^{\prime},v_i)$ with length 1 and attack cost $W+1$.
\item We set the fortification budget $B'$ to $B_2$, i.e. the value of the attack budget in UBIK, the attack budget $W'$ to $W$, i.e. the value of the weight capacity of UBIK, and the goal for the shortest path length to $\bar{K'}=\bar{K}+2n$. 
By construction, therefore, only the arcs $(v_{i-1},v_i^{\prime})$ can be attacked, and only within the capacity of the knapsack.
\end{itemize}
We provide a graphical view of the reduction in Figure~\ref{Fig:SPIPUF_Gadget}. Notice that the defender and the attacker adopt a reverse perspective in the two problems: fortifying arcs in SPIPUF corresponds to interdicting items in UBIK, and attacking arcs in SPIPUF to selecting items in UBIK. In fact, the rationale is to solve the KP by minimising the total profit of the items not selected. 

\begin{figure}[ht]
\begin{centering}
\begin{tikzpicture}[>=latex, semithick]

\node [circle, draw=white, semithick,  minimum size=1cm] (w1) at (3,3.3) {$p_1+2,W+1$};
\node [circle, draw=white, semithick,  minimum size=1cm] (w2) at (1.6,1.7) {$1,w_1$};
\node [circle, draw=white, semithick,  minimum size=1cm] (w3) at (4.7,1.7) {$1,W+1$};
\node [circle, draw=white, semithick,  minimum size=1cm] (w4) at (13,3.3) {$p_n+2,W+1$};
\node [circle, draw=white, semithick,  minimum size=1cm] (w5) at (11.6,1.7) {$1,w_n$};
\node [circle, draw=white, semithick,  minimum size=1cm] (w6) at (14.7,1.7) {$1,W+1$};

\node [circle, draw, semithick,  minimum size=0.6cm, label={[label distance=0.1cm]90:$v_0$}] (1) at (1,3) {};
\node [circle, draw, semithick,  minimum size=0.6cm, label={[label distance=0.1cm]-90:$v_1^{\prime}$}] (2) at (3,1) {};
\node [circle, draw, semithick,  minimum size=0.6cm, label={[label distance=0.1cm]90:$v_1$}] (3) at (5,3) {};
\node [circle, fill=black, draw, semithick,  minimum size=0.1cm, text width=0pt, text height=0pt, inner sep=0pt] (4) at (7,3) {};
\node [circle, fill=black, draw, semithick,  minimum size=0.1cm, text width=0pt, text height=0pt, inner sep=0pt] (5) at (8,3) {};
\node [circle, fill=black, draw, semithick,  minimum size=0.1cm, text width=0pt, text height=0pt, inner sep=0pt] (6) at (9,3) {};
\node [circle, draw, semithick,  minimum size=0.6cm, label={[label distance=0.1cm]90:$v_{n-1}$}] (7) at (11,3) {};
\node [circle, draw, semithick,  minimum size=0.6cm, label={[label distance=0.1cm]-90:$v_n^{\prime}$}] (8) at (13,1) {};
\node [circle, draw, semithick,  minimum size=0.6cm, label={[label distance=0.1cm]90:$v_n$}] (9) at (15,3) {};

\draw [->, semithick](1) -- (2);
\draw [->, semithick](1) -- (3);
\draw [->, semithick](2) -- (3);
\draw [->, semithick](7) -- (9);
\draw [->, semithick](7) -- (8);
\draw [->, semithick](8) -- (9);

\end{tikzpicture}
\end{centering}
\caption{Example of the graph obtained after reducing to SPIPUF. The label for each arc provides the length and the attack cost of the arc.}
\label{Fig:SPIPUF_Gadget}
\end{figure}

Let UBIK be a \emph{Yes} instance. We form a solution to SPIPUF as follows: if item $i\in I$ is interdicted by the attacker in UBIK, we fortify the arc $(v_{i-1},v_i^{\prime})$, after which no more fortification budget remains. Now the attacker in SPIPUF can attack only the unfortified arcs $(v_{i-1},v_i^{\prime})$ with $i=1,...,n$. For each set of attacked arcs $D$, the shortest path from $v_{0}$ to $v_{n}$ is easily determined: for each $i=1,...,n$, if arc $(v_{i-1},v_i^{\prime})$ is attacked, the defender has to use arc $(v_{i-1},v_i)$ with length $p_i+2$; otherwise, the defender will use arcs $(v_{i-1},v_i^{\prime})$ and $(v_i^{\prime},v_i)$ with total length 2. The worst case occurs when the attacker selects the attacked arcs $(v_{i-1},v_i^{\prime})$ so that the sum of the profits $p_i$ on the alternative arcs $(v_{i-1},v_i)$ is maximum. Since for each $i=1,...,n$ we either travel a length $2$ or $p_i+2$ (if an arc is interdicted in the associated gadget), the interdiction of an arc $(v_{i-1},v_i^{\prime})$ in SPIPUF corresponds to the selection of an item in UBIK. Since UBIK is a \emph{Yes} instance, we cannot achieve a profit of $\bar K$ or more; hence, the shortest path length cannot reach $\bar K+2n$. Therefore, the SPIPUF is a \emph{Yes} instance.

Conversely, assume that the SPIPUF is a \emph{Yes} instance. Only the arcs $(v_{i-1},v_i^{\prime})$ for $i=1,...,n$ are worth fortifying, since the other ones cannot be interdicted respecting the budget. Given a solution to SPIPUF, we can right away map the fortified arcs onto the corresponding items of UBIK. Since SPIPUF is a \emph{Yes} instance, the attacker cannot interdict a set of $(v_{i-1},v_i^{\prime})$ arcs that at the same time has a total attack budget not larger than $W$ and increases the shortest path length by at least $\bar K$. Mapping any set of interdicted arcs to the corresponding items in UBIK, we have that no item selection in UBIK can reach the profit goal $\bar K$ and UBIK is a \emph{Yes} instance, which completes the proof.
\end{proof}

\section{Multi-level Critical Node Problem with unit weights}
\label{sec:MCNP}

In this section, we first recall the formal definition of the MCNP with unit weights, which is defined in \cite{Baggio2021} as a three-level problem, after which we proceed to demonstrate its \Sth-completeness.

\paragraph{Multi-level Critical Node Problem}
The MCNP is defined on an undirected graph $G=(V,A)$. First, the defender selects a subset of vertices $D \subseteq V$ to \emph{vaccinate}, subject to a vaccination budget $\Omega$ and weight $\hat{c}_{v}$ for each $v \in V$. Then, the attacker selects a subset of vertices $I \subseteq V\setminus D$ to (directly) \emph{infect}, subject to an infection budget $\Phi$ and weight $h_v$ for $v \in V$. Finally, the defender selects a subset of vertices $P \subseteq V\setminus I$ to \emph{protect}, subject to a protection budget $\Lambda$ and weight $c_v$ for $v \in V$. A directly or indirectly infected vertex $v$ propagates the infection to a vertex $u$, if $(v,u) \in A$ and $u$ is neither vaccinated nor protected. The goal of the defender is to maximise the benefit $b_v$ of saved vertices (\ie, not infected), while the attacker aims to minimise it. We assume that all parameters of the problem are non-negative integers. 

The game description can be succinctly given by the following mixed integer trilevel program:

\begin{subequations}%
\begin{align}
(MCN)  \hspace{-0.4cm}
\max_{ \begin{subarray}{c}\\
 z \in \{0,1\}^{|V|} \\[1pt]
\displaystyle \sum_{v\in V} \hat{c}_v z_{v} \leq \Omega
 \end{subarray}} \quad
 & 
 \min_{  \begin{subarray}{c}\\ 
 y \in \{0,1\}^{|V|}\\[1pt] 
\displaystyle \sum_{v\in V} h_v y_{v} \leq \Phi
 \end{subarray}} \quad
 & 
 \max_{ 
 \begin{subarray}{c}\\ 
  x \in \{0,1\}^{|V|} \\[1pt] 
 \alpha \in \left[0,1\right]^{|V|}
 \end{subarray}
 } \ \ &\sum_{v\in V} b_v \alpha_{v} \nonumber\\
&  & s.t. \hspace*{0.7cm} & \hspace*{-0.4cm}   \sum_{v\in V} c_v x_{v}  \leq   \Lambda  \nonumber& \\
& &  &\hspace*{-0.4cm} \alpha_{v}  \leq 1 + z_{v} - y_{v}, &  \forall v\in V  \label{const:inf_vac} \\
& &  &\hspace*{-0.4cm} \alpha_{v}  \leq  \alpha_{u} + x_{v} + z_{v}, &  \ \ \forall\ (u,v)\in A, \label{const:proc_vac}
\end{align}
\label{MCN_formulation}%
\end{subequations}

where $z$, $y$, $x$ and $\alpha$ are binary decision vectors whose coordinates are $z_v$, $y_v$, $x_v$ and $\alpha_v$ for each $v \in V$. In this optimisation model, $z$, $y$ and $x$ reflect the set of vaccinated vertices $D=\lbrace v\in V: z_v=1 \rbrace$, directly infected vertices $I=\lbrace v\in V: y_v=1 \rbrace$ and protected vertices $P=\lbrace v\in V: x_v=1 \rbrace$, respectively. Finally, $\alpha$ mimics the propagation of the infection among the vertices in $V$, through Constraints~\eqref{const:inf_vac} and~\eqref{const:proc_vac}. Concretely, $\alpha_v=1$ means that vertex $v$ is saved and $\alpha_v=0$ means that vertex $v$ is infected. 
See~\cite{Baggio2021} for further details on this mathematical programming formulation and Figure~\ref{fig:example} for an illustration of the game.

\begin{figure}\centering
\resizebox{\textwidth}{!}{
\begin{tabular}{|c|c|c|c|}\hline
&&&\\
Instance & Vaccinate $v_3$ & Infect $v_2$ & Protect $v_1$\\
\begin{tikzpicture}[-,>=stealth',shorten >=0.2pt,auto,node distance=1cm,baseline={(current bounding box.north)},  scale=0.8, transform shape,
  main node/.style={circle,fill=black!15,draw,minimum size=0.7cm},spe node/.style={fill=white,draw=white}]
    
  \node[main node] (1) {$v_1$};
  \node[main node] (3) [below right=of 1] {$v_3$};
  \node[main node] (2) [above right =of 3] {$v_2$};
  \node[main node] (4) [below left=of 3] {$v_4$};
  \node[main node] (5) [below right=of 3] {$v_5$};
  \node[main node] (6)  [right= of 3] {$v_6$};

  \path[every node/.style={font=\sffamily\small}]
    (1) edge node {}  (4)
    (2) edge node {}  (1)
        edge node {}  (6)
    (3) edge node {}  (1)
        edge node {}  (2)
        edge node {}  (5)
    (4) edge node {}  (3)
    (5) edge node {}  (3)
     (6) edge node {}  (5);
\end{tikzpicture} &
\begin{tikzpicture}[-,>=stealth',shorten >=0.2pt,auto,node distance=1cm,baseline={(current bounding box.north)},  scale=0.8, transform shape,
  main node/.style={circle,fill=black!15,draw,minimum size=0.7cm},spe node/.style={fill=white,draw=white}]
  
  \node[main node] (1) {$v_1$};
  \node[main node] (3) [below right=of 1] {$v_3$};
  \node[main node] (2) [above right =of 3] {$v_2$};
  \node[main node] (4) [below left=of 3] {$v_4$};
  \node[main node] (5) [below right=of 3] {$v_5$};
  \node[main node] (6)  [right= of 3] {$v_6$};

  \path[every node/.style={font=\sffamily\small}]
    (1) edge node {}  (4)
    (2) edge node {}  (1)
        edge node {}  (6)
     (6) edge node {}  (5);
\end{tikzpicture}
&
\begin{tikzpicture}[-,>=stealth',shorten >=0.2pt,auto,node distance=1cm,baseline={(current bounding box.north)},  scale=0.8, transform shape,
  main node/.style={circle,fill=black!15,draw,minimum size=0.7cm},spe node/.style={circle,fill=black,draw,minimum size=0.7cm}]

  \node[main node] (1) {$v_1$};
  \node[main node] (3) [below right=of 1] {$v_3$};
  \node[spe node] (2) [above right =of 3] {\textcolor{white}{$v_2$}};
  \node[main node] (4) [below left=of 3] {$v_4$};
  \node[main node] (5) [below right=of 3] {$v_5$};
  \node[main node] (6)  [right= of 3] {$v_6$};

  \path[every node/.style={font=\sffamily\small}]
    (1) edge node {}  (4)
    (2) edge node {}  (1)
        edge node {}  (6)
    (6) edge node {}  (5);
\end{tikzpicture}
&
\begin{tikzpicture}[-,>=stealth',shorten >=0.2pt,auto,node distance=1cm,baseline={(current bounding box.north)},  scale=0.8, transform shape,
  main node/.style={circle,fill=black!15,draw,minimum size=0.7cm},spe node/.style={circle,fill=black,draw,minimum size=0.7cm}]

  \node[main node] (1) {$v_1$};
  \node[main node] (3) [below right=of 1] {$v_3$};
  \node[spe node] (2) [above right =of 3] {\textcolor{white}{$v_2$}};
  \node[main node] (4) [below left=of 3] {$v_4$};
  \node[spe node] (5) [below right=of 3] {\textcolor{white}{$v_5$}};
  \node[spe node] (6)  [right= of 3] {\textcolor{white}{$v_6$}};

  \path[every node/.style={font=\sffamily\small}]
    (2) edge node {}  (6)
     (6) edge node {}  (5);
\end{tikzpicture}\\ &&& \\ \hline
\end{tabular}
}
\caption{Example of an {\MCN} game with unitary weights and benefits, and budgets $\Omega=\Phi=\Lambda=1$. At the end of the game, vertices $\{v_1,v_3,v_4\}$ are saved and $\{v_2,v_6,v_5\}$ are infected.}
\label{fig:example}
\end{figure}

We recall that this problem was proved to be \Sth-complete on trees, with arbitrary integer weights, in \cite{NABLI2022122}.
In the following we will concentrate on the unweighted version, tackled numerically in \cite{Baggio2021}, for which all the weights previously described are unitary.

\paragraph{\Sth-completeness of the MCNP with unit weights}
We now demonstrate the completeness of the MCNP with unit weights for the third level of the polynomial hierarchy. This version of the MCNP is defined as:
\vspace{0.3cm}
\boxxx{
\textsc{\textbf{Unitary Multi-level Critical Node (UMCN)}}: \\
{\sc instance}: An undirected graph $G=(V,A)$, three positive integer maximum cardinalities $\Omega$, $\Phi$ and $\Lambda$ and a positive integer goal $\bar{K}$.\\
{\sc question}: Is there a subset $D\subseteq V$ of vertices, with $|D| \leq \Omega$, such that for every subset $I\subseteq V\setminus D$, with $|I| \leq \Phi$, there is a subset $P \subseteq V \setminus I$, with  $|P| \leq \Lambda$, such that the cardinality of the saved vertices is at least $\bar{K}$?
}
\vspace{0.3cm}
In order to prove the \Sth-completeness of the above problem, we will use $\mathcal B_3 \cap 3CNF$ as in Section~\ref{sec:Knapsack}.

Let us introduce the following reduction from the $\mathcal B_3 \cap 3CNF$:
\begin{itemize}
\item For each literal $\lambda = u$ or $\lambda = \bar{u}$ with $u \in U$, we introduce in the graph a vertex $v_{\lambda}$. We also define $V_X=\{v_x,v_{\bar x}: x\in X\}$, $V_Y=\{v_y,v_{\bar y}: y\in Y\}$ and $V_Z=\{v_z,v_{\bar z}: z\in Z\}$.
\item For each 
unordered pair of literals $\{ \lambda,\lambda' \}$ with $\lambda, \lambda' \in \{ x,\bar{x} \}$, $\lambda \neq \lambda'$ and $x \in X$, we add a gadget (see the left part of Figure~\ref{Fig:Gadget}), composed by a set $C_{\lambda\lambda'}$ of $\Gamma_X$ vertices, all connected with $v_{\lambda}$ and $v_{\lambda'}$. The cardinality $\Gamma_X$ is suitably defined in the following.
The aim of these gadgets is to turn each couple of vertices from $V_X$ into a vertex cut, so that protecting both vertices allows to save $C_{\lambda\lambda'}$.
Since this does not hold for the couple of vertices $(v_x,v_{\bar x})$ for each $x\in X$, it will be effective to protect only one vertex from such couples.
\item For each 
literal $\lambda = y$ or $\lambda = \bar{y}$ with $y\in Y$, we introduce a set $S_{\lambda}$ of $\Gamma_Y$ vertices, each with an edge linking it to $v_{\lambda}$. Therefore, $v_y$ and $v_{\bar y}$ are at the centre of a substar inside the graph. 
We also introduce a set $K_y$ of $\Gamma_Y$ vertices, all linked to both $v_y$ and $v_{\bar y}$. The value of $\Gamma_Y$ is defined in the following.
We provide a graphic illustration of such a gadget in the central part of Figure~\ref{Fig:Gadget}.
\item We add edges between each vertex of $V_X\cup V_Z$ and each vertex of $V_Y$.
\item For each 
unordered pair of literals $\{ \lambda,\lambda' \}$ with $\lambda, \lambda' \in \{ z,\bar{z} \}$ with $z \in Z$, $\lambda \neq \lambda'$ and $\lambda \neq \bar\lambda'$, we introduce a set $C_{\lambda\lambda'}$ of $\Gamma_Z$ vertices, all connected with $v_{\lambda}$ and $v_{\lambda'}$. The cardinality $\Gamma_Z$ is suitably defined in the following. 
The aim of this set of gadgets is once again to allow each couple of vertices from $V_Z$ to be part of a vertex cut, so that protecting both vertices allows to save $C_{\lambda\lambda'}$.
Since this does not hold for the couple of vertices $(v_z,v_{\bar z})$ for each $z\in Z$, it will be effective to protect only one vertex from such couples.
\item  For each clause $c \in C$, let $k_c \leq 3$ be the number of
literals included in the clause. We enumerate the $2^{k_c}-1$ variable assignments that satisfy the clause (excluding only the one which negates all literals) and introduce a vertex $v_c^i$ (with $i = 1,\ldots,2^{k_c}-1$) for each corresponding set of literals. We define as $V_C$ the set of all such vertices. These vertices are connected by edges to the corresponding vertices in $V_{X}$, $V_{Y}$ and $V_{Z}$.

The aim of these gadgets is to guarantee that exactly one of the vertices in each clause gadget will be saved if at least one of the vertices associated to the literals is vaccinated or protected, and no vertex is saved otherwise. We provide a graphic illustration of such a gadget in the right part of Figure~\ref{Fig:Gadget}.
\item We set the vaccination budget as $\Omega=|X|$, the infection budget as $\Phi=|X|+|Y|$ and the protection budget as $\Lambda=|Y|+|Z|$. We set the profit goal at $\bar K=\Gamma_X|X|(|X|-1)/2+|X|+|Y|(\Gamma_Y+1)+\Gamma_Z|Z|(|Z|-1)/2+|Z|+|C|$. Finally, we set $\Gamma_Z=|C|+1$, $\Gamma_Y=\Gamma_Z|Z|(|Z|-1)/2+1$ and $\Gamma_X=4|Y|\Gamma_Y+1$. This guarantees that it is more advantageous to save or infect (from the point of view of the defender or the attacker, respectively) vertices from $V_X$ first, from $V_Y$ second, and lastly from $V_Z$.
\end{itemize}

\begin{figure}[ht]
\begin{centering}
\include{Gadget}
\end{centering}
\caption{On the left part: example of a gadget between the vertices associated to two variables $x,x'\in X$ (which forms a bi-partite graph). The black dots denote the rest of the sets elliptically. On the middle part: example of a gadget associated to a variable $y\in Y$. The black dots denote the rest of the sets elliptically. On the right part: example of a gadget associated to a clause $c=x\land y\land\neg z$ with $x\in X$, $y\in Y$ and $z\in Z$. The only missing configuration on the right is the one of the negation of the clause, i.e., $\neg x\land\neg y\land z$, therefore in the case this expression is true in the SAT instance, no vertex on the right is saved from infection.}
\label{Fig:Gadget}
\end{figure}
The reduction is polynomial because the number of vertices and edges of the graph is polynomial in the instance size.

We proceed to prove some useful properties about the solutions to MCNP instances created above:
\begin{prop}\label{prop:MCNP1}
    In every \emph{Yes} instance of UMCN, the defender must vaccinate either vertex $v_x$ or $v_{\bar x}$ for each $x\in X$ in order to reach the profit goal. 
\end{prop}
\begin{proof}
    First of all, notice that, once a vertex $v_{\lambda} \in V_{X} \cup V_{Y} \cup V_{Z}$ is infected, the only way for the defender to save the vertices from the sets $C_{\lambda\lambda'}$ or $S_{\lambda}$ or the ones in $V_{C}$ that correspond to truth assignments using $\lambda$ is to vaccinate or protect them individually. This is ineffective when compared to the vaccination or protection of vertices from $V_X$, $V_Y$ and $V_Z$, that indirectly save also the adjacent vertices. Second, with the given $\Gamma$ parameters and in order to reach the profit goal, the defender must save at least $|X|(|X|-1)/2$ subsets $C_{\lambda\lambda'}$ of $\Gamma_X$ vertices by spending the whole vaccination budget on $V_X$. Otherwise, the attacker can infect $|X|+1$ vertices from $V_X$ directly, and the corresponding subsets $C_{\lambda\lambda'}$ indirectly.     
    Moreover, if the defender does not vaccinate exactly one vertex between $v_x$ and $v_{\bar x}$ for each $x\in X$, then some couples are completely vaccinated while others have both vertices unvaccinated, and therefore prone to infection. By construction, since the pairs $(v_x,v_{\bar x})$ do not have a corresponding $C$ subset, the number of saved subsets $C_{\lambda\lambda'}$ is strictly less than $|X|(|X|-1)/2$ and the profit goal $\bar K$ cannot be reached.
\end{proof}
\begin{prop}\label{prop:MCNP2}
    If the defender vaccinates exactly one vertex for each pair in $V_{X}$ and the attacker does not infect exactly one vertex for each pair in $V_{Y}$, the defender can always reach the profit goal.
\end{prop}
\begin{proof}
    If the defender vaccinates exactly one vertex for each pair in $V_{X}$, the attacker must infect all unvaccinated $V_X$ vertices. Otherwise, the defender will save at least $\Gamma_X|X|(|X|+1)/2>\bar K$ vertices.
    With the remaining $|Y|$ units of attack, infecting a vertex 
    from the $C_{\lambda\lambda'}$ set for literals $\lambda$ and $\lambda'$ of $X$ only infects this single vertex (a similar reasoning applies to the vertices of the $S_\lambda$ and $K_y$ sets, as well as the $C_{\lambda\lambda'}$ sets for $\lambda$ and $\lambda'$ being $Z$-related literals). Moreover, this attack would leave a remaining budget of $|Y|-1$ vertices. 
    From the remaining vertices of the graph, those of $V_Y$ are the ones with the largest number of adjacent nodes that can easily be saved by protecting a vertex in $V_Y$. Consequently, the defender can use $|Y|+1$ units from the protection budget to protect $|Y|+1$ vertices from $V_Y$ to obtain a number of saved vertices greater than $\Gamma_X|X|(|X|-1)/2+|X|+\Gamma_Y(|Y|+2)>\bar K$.

    Moreover, the attacker must infect either vertex $v_y$ or $v_{\bar y}$ for each $y\in Y$. Otherwise, there exists $y \in Y$ such that neither $v_y$ nor $v_{\bar y}$ are directly infected. Since the defender has enough budget to protect the $|Y|$ vertices of $V_Y$ not infected, both $v_y$ and $v_{\bar y}$ can be protected, saving the stars around the two vertices and the set $K_y$ linked to both vertices. This generates a number of saved vertices larger than $\Gamma_X|X|(|X|-1)/2+|X|+\Gamma_Y(|Y|+2)>\bar K$, which completes the proof.
\end{proof}
\begin{prop}\label{prop:MCNP3}
    If the defender vaccinates either vertex $v_x$ or $v_{\bar x}$ for each $x\in X$, and the attacker infects either vertex $v_y$ or $v_{\bar y}$ for each $y\in Y$,
    the defender can reach the profit goal only by protecting all the uninfected vertices from $V_Y$ and either vertex $v_z$ or $v_{\bar z}$ for each $z\in Z$.
\end{prop}
\begin{proof}
    Consider a vaccination and infection solution as described in the statement above. The defender must protect all the non-infected vertices of $V_Y$, otherwise the profit will be $\leq \Gamma_X|X|(|X|-1)/2+|X|+\Gamma_Y(|Y|+1)<\bar K$. With the remaining protection budget of $|Z|$ units, defending vertices from the subsets $C$, $S$ or $K$ subsets that are linked to an infected vertex is unfavourable, since doing so only saves the protected vertex. The defender would then have less than $|Z|$ remaining units to protect vertices from $V_Z$ and the number of saved vertices will be less than $\Gamma_X|X|(|X|-1)/2+|X|+\Gamma_Y(|Y|+1)+\Gamma_Z(|Z|-1)(|Z|-2)/2+|Z|+|C|<\bar K$.

    The defender will then protect exactly $|Z|$ vertices from set $V_Z$. Following the reasoning of Property~\ref{prop:MCNP1}, protecting both vertices $v_z$ and $v_{\bar z}$ for a given $z\in Z$ is ineffective.
    Therefore, the defender will protect exactly one $v$ vertex for each $z\in Z$. 
\end{proof}
\vspace{0.3cm}
\begin{thm}
UMCN is \Sth-complete.
\end{thm}
\begin{proof}
Given the form of the problem, UMCN is straightforwardly in \Sth.

Let $\mathcal B_3 \cap 3CNF$ be a \emph{Yes} instance. We form a solution to UMCN by selecting in $D$ the vertices $v_x$ such that $x \in X$ is 1 and the vertices $v_{\bar{x}}$ otherwise. Given Properties \ref{prop:MCNP2} and \ref{prop:MCNP3}, the attacker's solutions which do not make the profit $\bar K$ trivially reachable by the defender are those that infect exactly one vertex between $v_y$ and $v_{\bar y}$ for each $y\in Y$. Moreover, the defender's solutions that allow the defender to reach profit $\bar K$ are those for which the defender protects either vertex $v_z$ or $v_{\bar z}$ for each $z\in Z$. Therefore, any variable assignment of $\mathcal B_3 \cap 3CNF$ can be mapped to a solution of UMCN which does not automatically reach the profit goal: for any $y\in Y$ assigned a value 0 (and only those), assign $v_y$ to infection set $I$ (i.e., infect it) and then let the defender protect the remaining vertices of $V_Y$; for any $z\in Z$ assigned value 1 (and only those), assign $v_z$ to protection set $P$. Since the $\mathcal B_3 \cap 3CNF$ instance is a \emph{Yes} instance, for each clause, at least one of the vertices corresponding to the literals is vaccinated or protected. Therefore, one of the vertices corresponding to the clause is saved (the one connected by an edge to the three vertices associated to the literal which is true in the SAT instance), which guarantees that the defender can reach the profit goal $\bar K$ and UMCN is a \emph{Yes} instance.

Next, suppose that UMCN is a \emph{Yes} instance. According to Property \ref{prop:MCNP1}, the set $D$ must contain exactly one of the vertices $v_x$ and $v_{\bar{x}}$ for $x \in X$. Given such a set $D$, we can construct a solution for $\mathcal B_3 \cap 3CNF$ where, for each $x\in X$, we assign 1 to $x$ if $v_x\in D$ and 0 otherwise. For each assignment of the variables of set $Y$ in the SAT instance, we consider an attacker's solution where for each $y\in Y$, we assign 0 to $y$ if $y\in I$ and 1 otherwise. Given Property \ref{prop:MCNP3}, each lower level reaction of the defender protects exactly one vertex between $v_z$ and $v_{\bar z}$ for each $z\in Z$. Therefore we can map this solution to a SAT solution where for each $z\in Z$, we assign 1 to $z$ if $v_z\in P$ and 0 otherwise. Since UMCN is a \emph{Yes} instance, in addition to the set of $\Gamma_X|X|(|X|-1)/2+|X|+|Y|(\Gamma_Y+1)+\Gamma_Z|Z|(|Z|-1)/2+|Z|$ vertices saved directly by the defender's selection, it needs to save one vertex per clause of $C$ (and exactly that amount since for each clause, a maximum of one vertex can be saved). Therefore, one of the literals of each clause needs to be assigned the value 1, ensuring that the $\mathcal B_3 \cap 3CNF$ is a \emph{Yes} instance, which completes the proof.
\end{proof}

Compared to the proof proposed in \cite{NABLI2022122}, which reduced the TIKP to the MCNP over trees, the reduction proposed here is based on graphs with a more complex topology. This topology is dictated, like the reduction presented earlier for the Interdiction Knapsack, by the need to make the nodes from, e.g., $V_X$ more interesting to focus on for both players. This type of construction is recurrent in all our proofs relying on multi-level SAT problems, due to the symmetry of the decisions for the higher levels because of the equality of all fortification and attack costs, as can be seen also in the reduction used to prove the \Stwo-completeness of the MFIPF in the previous section.

\section{Conclusion}
\label{sec:conclusion}

Over the years, the literature about interdiction problems with fortification has built slowly but steadily, yet until recently~\cite{NABLI2022122}, no completeness result for such tri-level problems was rigorously proved. In this work, we provided proofs clarifying the position of several fortification problems in the polynomial hierarchy. Since such problems are usually computationally tackled with unit fortification and attack costs, we have provided reductions that assume unit fortification costs and, whenever possible, also unit attack costs, refining on the previous results of~\cite{NABLI2022122}. We investigated problems whose structure may prove useful in the future in clarifying the complexity of other fortification problems. For example, we considered the Multi-level Interdiction Knapsack Problem, which we proved to be complete for any level of the polynomial hierarchy when one allows for an arbitrary number of fortification-interdiction rounds, or the Max-Flow Interdiction Problem with Fortification, which we proved to be \Stwo-complete. Using our result on the Bi-level Interdiction Knapsack Problem, we have proved that some well-studied fortification problems, such as the Shortest Path Interdiction Problem with Fortification or a well-known Tri-level Electric Power Grid Fortification Problem, are \Stwo-complete. Finally, we refined a result of~\cite{NABLI2022122} by showing that the Multi-level Critical Node Problem is \Sth-complete even with unit weights and profits.

Further investigation on the computational complexity of these problems can take several forms. The study of the Multi-level Interdiction KP is interesting as the problem does not have much structure, differently from a network interdiction-fortification problem. This fact makes it easier to use it for deriving the completeness of other interdiction-fortification problems, similar to what is done in Section~\ref{sec:ShortestPath} and Appendix~\ref{sec:PowerGrid}. Moreover, it is interesting that even with unit fortification and interdiction costs, the problem complexity remains the same, as long as the lowest level weights and profits are arbitrary, whereas if either these weights or profits become unitary, no matter the number of interdiction-fortification rounds, the problem collapses to P. An interesting avenue of research would be to try to find out a special structure of the lowest level weights and/or profits that would decrease the problem complexity without rendering it polynomial. Alternatively, if one of the lowest level sets of weights and profits becomes unitary (so that the lowest level KP becomes polynomial), what is the complexity of the problem when at least one set of higher level costs is instead arbitrary? Such clarifications may help to understand the articulation of the different levels which defines the intrinsic complexity of higher level problems in the polynomial hierarchy. Another interesting subject is the status of specially structured instances of problems formulated on undirected graphs, such as the Multi-level Critical Node Problem with unit weights and profits: while the weighted counterpart has been proved to be \Sth-complete even for the simple case of trees, the reduction used in this work for the unweighted case uses far more structured graphs. Since the lower level of this problem becomes polynomial over trees~\cite{NABLI2022122}, it would be interesting to investigate the full unweighted tri-level problem on specially structured graphs. Finally, exploring the complexity of other existing fortification problems or clarifying the case of the SPIPF or TEPGFP with unit fortification and attack costs could be a worthy goal. We also underline that other interdiction-protection problems exist which do not rely strictly on fortification as a protection mechanism, see e.g. \cite{Barbosa2021}, and that the results obtained in this work may be of help in the study of their computational complexity.

\bibliographystyle{plain} 
\bibliography{Biblio}

\begin{thebibliography}{10}

\bibitem{Addis2013}
Bernardetta Addis, Marco Di~Summa, and Andrea Grosso.
\newblock Identifying critical nodes in undirected graphs: Complexity results
  and polynomial algorithms for the case of bounded treewidth.
\newblock {\em Discrete Applied Mathematics}, 161(16):2349--2360, Nov 2013.

\bibitem{ARULSELVAN20092193}
Ashwin Arulselvan, Clayton~W. Commander, Lily Elefteriadou, and Panos~M.
  Pardalos.
\newblock Detecting critical nodes in sparse graphs.
\newblock {\em Computers \& Operations Research}, 36(7):2193--2200, 2009.

\bibitem{assimakopoulos}
Nikitas Assimakopoulos.
\newblock A network interdiction model for hospital infection control.
\newblock {\em Computers in Biology and Medicine}, 17(6):413--422, 1987.

\bibitem{Baggio2021}
Andrea Baggio, Margarida Carvalho, Andrea Lodi, and Andrea Tramontani.
\newblock Multilevel approaches for the critical node problem.
\newblock {\em Operations Research}, 69:486--508, 2021.

\bibitem{Barbosa2021}
Fábio Barbosa, Agostinho Agra, and Amaro {de Sousa}.
\newblock The minimum cost network upgrade problem with maximum robustness to
  multiple node failures.
\newblock {\em Computers \& Operations Research}, 136:105453, 2021.

\bibitem{Brown2006}
Gerald Brown, Matt Carlyle, Javier Salmerón, and R.Kevin Wood.
\newblock Defending critical infrastructure.
\newblock {\em Interfaces}, 36:530--544, 12 2006.

\bibitem{Caprara2016}
A.~Caprara, M.~Carvalho, A.~Lodi, and G.~J. Woeginger.
\newblock Bilevel knapsack with interdiction constraints.
\newblock {\em INFORMS Journal on Computing}, 28:319--333, 2016.

\bibitem{Caprara2014}
Alberto Caprara, Margarida Carvalho, Andrea Lodi, and Gerhard~J. Woeginger.
\newblock A study on the computational complexity of the bilevel knapsack
  problem.
\newblock {\em SIAM Journal of Optimization}, 24:823--838, 2014.

\bibitem{Chen2023}
F.~Chen, R.~Wang, Z.~Xu, H.~Liu, and M.~Wang.
\newblock A quantified multi-stage optimization method for resource allocation
  of electric grid defense planning.
\newblock {\em Electric Power Systems Research}, 220:109284, 2023.

\bibitem{IntrAlg2009}
Thomas~H. Cormen, Charles~E. Leiserson, Ronald~L. Rivest, and Clifford Stein.
\newblock {\em Introduction to Algorithms, Third Edition}.
\newblock The MIT Press, 3rd edition, 2009.

\bibitem{Cormican1998}
K.~J. Cormican, D.~P. Morton, and R.~K. Wood.
\newblock Stochastic network interdiction.
\newblock {\em Operations Research}, 46:184--197, 1998.

\bibitem{Delgadillo2010}
A.~Delgadillo, J.~M. Arroyo, and N.~Alguacil.
\newblock Analysis of electric grid interdiction with line switching.
\newblock {\em IEEE Transactions on Power Systems}, 25:633--641, 2010.

\bibitem{Croce2020A}
Federico {Della Croce} and Rosario Scatamacchia.
\newblock An exact approach for the bilevel knapsack problem with interdiction
  constraints and extensions.
\newblock {\em Mathematical Programming}, 183:249--281, 2020.

\bibitem{DeNegre2011}
Scott DeNegre.
\newblock {\em Interdiction and Discrete Bilevel Linear Programming}.
\newblock PhD thesis, Lehigh University, 2011.

\bibitem{Enayaty-Ahangar2019}
F.~Enayaty-Ahangar, C.~E. Rainwater, and T.~C. Sharkey.
\newblock A logic-based decomposition approach for multi-period network
  interdiction models.
\newblock {\em Omega}, 87:71--85, 2019.

\bibitem{Fakhry2022}
Ramy Fakhri, Elkafi Hassini, Mohamed Ezzeldin, and Wael El-Dakhakhni.
\newblock Tri-level mixed-binary linear programming: Solution approaches and
  application in defending critical infrastructure.
\newblock {\em European Journal of Operational Research}, 298:1114--1131, 2022.

\bibitem{Frohlich2021}
Nicolas Fr\"ohlich and Stefan Ruzika.
\newblock On the hardness of covering-interdiction problems.
\newblock {\em Theoretical Computer Science}, 871:1--15, 2021.

\bibitem{Furini2019}
Fabio Furini, Ivana Ljubic, Sebastien Martin, and Pablo San~Segundo.
\newblock The maximum clique interdiction problem.
\newblock {\em European Journal of Operational Research}, 277:112--127, 02
  2019.

\bibitem{Goerigk2024}
M.~Goerigk, S.~Lendl, and L.~Wulf.
\newblock On the complexity of robust multi-stage problems with discrete
  recourse.
\newblock {\em Discrete Applied Mathematics}, 343:355--370, 2024.

\bibitem{Israeli2002}
E.~Israeli and R.~K. Wood.
\newblock Shortest-path network interdiction.
\newblock {\em Networks}, 40(2):97--111, 2002.

\bibitem{Johannes2011NewCO}
Berit Johannes.
\newblock {\em New Classes of Complete Problems for the Second Level of the
  Polynomial Hierarchy}.
\newblock PhD thesis, Technischen Universitat Berlin, 2011.

\bibitem{Kuttler2024}
E.~Kuttler, N.~Ghorbani-Renani, K.~Barker, and A.~D. Gonz\'alez.
\newblock Protection-interdiction-restoration for resilient multi-commodity
  networks.
\newblock {\em Reliability Engineering and System Safety}, 242:109745, 2024.

\bibitem{LALOU201892}
Mohammed Lalou, Mohammed~Amin Tahraoui, and Hamamache Kheddouci.
\newblock The critical node detection problem in networks: A survey.
\newblock {\em Computer Science Review}, 28:92--117, 2018.

\bibitem{LEITNER20231026}
Markus Leitner, Ivana Ljubić, Michele Monaci, Markus Sinnl, and Kübra
  Tanınmış.
\newblock An exact method for binary fortification games.
\newblock {\em European Journal of Operational Research}, 307(3):1026--1039,
  2023.

\bibitem{Lozano2017b}
L.~Lozano, J.~C. Smith, and M.~E. Kurz.
\newblock The traveling salesman problem with interdiction and fortification.
\newblock {\em Annals of Operations Research}, 45:210--216, 2017.

\bibitem{Lozano2017}
Leonardo Lozano and J.~Cole Smith.
\newblock A backward sampling framework for interdiction problems with
  fortification.
\newblock {\em INFORMS Journal on Computing}, 29:123--139, 2017.

\bibitem{malaviya}
Ajay Malaviya, Chase Rainwater, and Thomas Sharkey.
\newblock Multi-period network interdiction problems with applications to
  city-level drug enforcement.
\newblock {\em IIE Transactions}, 44(5):368--380, 2012.

\bibitem{NABLI2022122}
Adel Nabli, Margarida Carvalho, and Pierre Hosteins.
\newblock Complexity of the multilevel critical node problem.
\newblock {\em Journal of Computer and System Sciences}, 127:122--145, 2022.

\bibitem{Nguyen2024}
D.~H. Nguyen and J.~C. Smith.
\newblock Asymmetric stochastic shortest-path interdiction under conditional
  value-at-risk.
\newblock {\em IISE Transactions}, 56:398--410, 2024.

\bibitem{Rutenburg1994}
V.~Rutenburg.
\newblock Propositional truth maintenance systems: Classification and
  complexity analysis.
\newblock {\em Annals of Mathematics and Artificial Intelligence}, 10:207--231,
  1994.

\bibitem{Salmeron2004}
J.~Salmeron, K.~Wood, and R.~Baldick.
\newblock Analysis of electric grid security under terrorist threat.
\newblock {\em IEEE Transactions on Power Systems}, 19:905--912, 2004.

\bibitem{Salmeron2009}
J.~Salmeron, K.~Wood, and R.~Baldick.
\newblock Worst-case interdiction analysis of large-scale electric power grids.
\newblock {\em IEEE Transactions on Power Systems}, 24:96--104, 2009.

\bibitem{Stockmeyer1976}
Larry~J. Stockmeyer.
\newblock The polynomial-time hierarchy.
\newblock {\em Theoretical Computer Science}, 3(1):1--22, Oct 1976.

\bibitem{Sullivan2014}
K.~M. Sullivan and J.~C. Smith.
\newblock Exact algorithms for solving a euclidean maximum flow network
  interdiction problem.
\newblock {\em Navigation, Journal of the Institute of Navigation},
  64:109--124, 2014.

\bibitem{Sullivan2014b}
Kelly~M. Sullivan, David~P. Morton, Feng Pan, and J.~Cole~Smith.
\newblock Securing a border under asymmetric information.
\newblock {\em Naval Research Logistics (NRL)}, 61(2):91--100, 2014.

\bibitem{Weninger2023}
Noah Weninger and Ricardo Fukasawa.
\newblock A fast combinatorial algorithm for the bilevel knapsack problem with
  interdiction constraints.
\newblock In Alberto Del~Pia and Volker Kaibel, editors, {\em Integer
  Programming and Combinatorial Optimization}, pages 438--452, Cham, 2023.
  Springer International Publishing.

\bibitem{Wood1993}
R.~Kevin Wood.
\newblock Deterministic network interdiction.
\newblock {\em Mathematical and Computer Modelling}, 17(2):1--18, Jan 1993.

\bibitem{Wrathall1976}
Celia Wrathall.
\newblock Complete sets and the polynomial-time hierarchy.
\newblock {\em Theoretical Computer Science}, 3(1):23--33, Oct 1976.

\bibitem{Wu2017}
Xuan Wu and Antonio~J. Conejo.
\newblock An efficient tri-level optimization model for electric grid defense
  planning.
\newblock {\em IEEE Transactions on Power Systems}, 32:2984--2994, 2017.

\bibitem{Wu2021}
Yipeng Wu, Zhilong Chen, Huadong Gong, Qilin Feng, Yicun Chen, and Haizhou
  Tang.
\newblock Defender-attacker-operator: Tri-level game-theoretic interdiction
  analysis of urban water distribution networks.
\newblock {\em Reliability Engineering and System Safety}, 214:107703, 2021.

\end{thebibliography}

\begin{appendices}
\section{Proof the \Sth{}-completeness of the Unitary Multi-level Interdiction Knapsack}\label{app:UMIKP}
We provide here the proof of Theorem \ref{thm:UMIK}.
\begin{proof}
Starting from an instance of $\mathcal B_{m+1} \cap \overline{3CNF}$ (if $m$ odd), or $\mathcal B_{m+1} \cap 3CNF$ (if $m$ even), we construct an instance of UMIK as follows:
\begin{itemize}
    \item For each variable $u \in U$, we create two items $i_u$ and $i_{\bar{u}}$, one for each possible 0-1 assignment of $u$. We define $I_{X_l} = \lbrace i_x,i_{\bar x}: x \in X_l \rbrace$ with $l=1,\ldots,m+1$. For variables of each interdiction level $x\in X_l$ with $l$ an even number, we introduce a third item $j_x$ and a fourth item $j_x^\prime$, which will serve to force the attacker to choose one and only one item between $i_x$ and $i_{\bar{x}}$ and the defender to consume the knapsack capacity completely.
    \item For each clause $c \in C$, we create two items $i^1_c$, $i^{2}_c$. We designate by $I_C$ the set of items associated with $C$.
    \item Weights, profits, maximum capacities, maximum profit and goal will be given by positive integer numbers expressed in base 10.
    Each digit position is labelled by a variable or a clause: the first $\vert C \vert$ positions (least significant numbers) are labelled by the clauses; the next $\vert X_1 \vert$ positions are labelled by the variables $X_1$, the next $2|X_2|$ positions are labelled by the variables $X_2$, etc... until the next $2\vert X_{m+1} \vert$ positions, if $m$ is odd, or $\vert X_{m+1} \vert$ positions, if $m$ is even, which are labelled by the variables $X_{m+1}$; finally, the last (highest) positions are labelled from $M$ on.
        \begin{itemize}
            \item All items have attack and fortification costs equal to 1 (\ie, unitary costs).
            \item For each level with an odd index $l$, that is for all fortification levels and for the regular KP on the lowest level, the two items $i_x$ and $i_{\bar{x}}$ corresponding to each variable $x \in X_l$ have weights and profits 
            $w_{i_x}$, $p_{i_x}$, $w_{i_{\bar{x}}}$ and $p_{i_{\bar{x}}}$ 
            with digit 1 in the position labelled by the variable $x$ and 0 in the positions labelled by other variables.
            
            If literal $x$ appears in clause $c \in C$, then $p_{i_x}$ and $w_{i_x}$ have digit 1 in the position labelled as $c$, and 0 otherwise. Similarly, if literal $\neg x$ appears in clause $c \in C$,  $p_{i_{\bar{x}}}$ and $w_{i_{\bar{x}}}$ have digit 1 in the position labelled by $c$, and 0 otherwise.

            Finally, if $l\geq3$, $\forall x\in X_l$, $p_{i_x}$ and $p_{i_{\bar x}}$ have value $\delta_l=\prod_{l' = 2}^{l-1}(|X_{l'}|+1)$ encoded in base 10 in the positions from $M$ on.
            
            \item For each interdiction level $l$ (with $l$ an even number), the two items $i_x$ and $i_{\bar{x}}$ corresponding to $x \in X_l$ have weights and profits with digit 1 in the higher position labelled by variable $x$ and 0 in the lower position labelled by $x$ and in the positions labelled by other variables.
            
            The profits $p_{i_x}$ and $p_{i_{\bar{x}}}$ have, $\forall x\in X_l$, in the positions from $M$ on, value $\delta_l=\prod_{l' = 2}^{l-1}(|X_{l'}|+1)$  
            if $l \geq 4$, and value $\delta_2=1$ if $l=2$. If literal $x$ appears in clause $c \in C$, then the profit and weight have digit 1 in the position labelled as $c$, and 0 otherwise. Similarly, if literal $\neg x$ appears in clause $c \in C$, the profit and weight have digit 1 in the position labelled by $c$, and 0 otherwise.
            
            The item $j_x$ corresponding to $x$ has a weight $w_{j_x}$ and profit $p_{j_x}$ with digit 2 in the higher position labelled as $x$ and 0 elsewhere.

            Finally, the item $j_x^\prime$ corresponding to $x$ has a weight $w_{j_x^\prime}$ with digit 1 in the higher position labelled as $x$ and 0 elsewhere and a profit $p_{j_x^\prime}$ with digit 1 in the lower position labelled as $x$ and 0 elsewhere.
            \item For each $c \in C$, the first item $i^1_c$ has weight and profit with digit 1 in the position labelled as $c$ and 0 elsewhere, and the second item $i^2_c$ has weight and profit with digit 2 in the position labelled as $c$ and 0 elsewhere.
            \item The attack budget for even levels is $B_l=|X_l|$, while the fortification budget for odd levels $l\geq3$ is $B_l=|X_l|+|X_{l+1}|$ if $l$ is not the highest decision level and $B_l=|X_l|$ otherwise.
            \item The capacity $W$ has 1s for all digits with labels associated to either the fortification levels or the lowest level, 2s for all digits corresponding to the higher position associated with labels in interdiction levels, 0s in the lower position associated with each label in interdiction levels and 4s for all digits with labels in $C$, and 0 for digit $M$.
            \item The profit goal $\bar{K}$ has digit 1 in all positions with labels in $U$ and 4 in all positions with labels in $C$, and value $\sum_{l=2}^{m+1}\delta_l|X_l|$ in the positions starting at $M$.
        \end{itemize}
\end{itemize}
The reduction is still polynomial because the profit goal $\bar{K}$ has $\vert C\vert +\sum_{k=1}^{m+1}\left(1+\frac{1+(-1)^k}{2}\right)\vert X_k \vert +\lceil\log_{10}(\sum_{l=2}^{m+1}\delta_l|X_l|)\rceil$ digits and all other numbers are smaller.
The proof of completeness is based mainly on the following properties:
\begin{itemize}
    \item The defender must select exactly one item for each pair associated with a variable at each level $I_{X_l}$ for $l=1,...,m+1$, plus all the $j_x^\prime$ items defined in the interdiction levels; this is allowed by the capacity and the fortification budgets.
    \item At any fortification level $l$ ($l$ odd), the defender fortifies the uninterdicted items corresponding to the above decision level of the 3-SAT problem, as well as one item for each pair of $I_{X_l}$, due to the hierarchical nature of the profits. The structure of the weights at the lowest level compels
    to fortify only one item between $i_x$ and $i_{\bar x}$ for a given $x\in X_l$, as the defender cannot select both such items together.
    \item At any interdiction level $l$ ($l$ even), the attacker interdicts one item for each pair of $I_{X_l}$, due to the hierarchical nature of the profits. If the attacker does not interdict one and only one item between $i_x$ and $i_{\bar x}$, some item $j_{x'}$ at some decision level can be selected by the defender in the last round of decision, which automatically allows the defender to reach $\bar K$.
    \item The fact that the defender can or cannot reach a value of 4 for each digit labelled in $C$ of both profit and weight of the lowest level is equivalent to fulfilling or not the 3-SAT formula.
\end{itemize}
Since the logic of the proof is very similar to the ones of the completeness of UBIK and UTIK developed above, we omit further details. We only observe that reasonings similar to the previous proofs will lead the defender to saturate the capacity at each digit, leaving no residual capacity to select items that correspond to lower decision levels of the SAT instance. 
\end{proof}

\section{The Tri-level Electric Power Grid Fortification Problem}
\label{sec:PowerGrid}

The last problem we will tackle is a power grid fortification-interdiction problem. The interdiction of electric power grids is a well studied interdiction problem \cite{Delgadillo2010,Salmeron2004,Salmeron2009}, and the same holds for its fortification version \cite{Chen2023,Fakhry2022,Wu2017}.

\paragraph{Problem definition}
The problem requires to decide the amount of power generated and routed through a distribution network in order to feed a set of demand points. In its interdiction version, the attacker disables elements of the power network, subjected to an attack budget, to prevent the defender from covering the whole demand. The protection version adds a further top decision layer in which the defender fortifies some of the network elements, making them impossible to disable. We will limit ourselves to the case where the attacks concentrate on the elements that are easier to physically access, i.e., the transmission lines.

The model of the power grid considers a set $J$ of generators, 
a set $L$ of transmission lines and a set $N$ of buses, formally defined as the set of start and end points of the transmission lines. Each generator or demand point is attached to a specific bus. The system can be seen as a graph where each node $n$ models a bus, each arc models the transmission line $l$ between an origin bus $O_l$ and a destination bus $D_l$. The subset of generators $J_n \subseteq J$ connected to bus $n\in N$ acts as a source of power flow. The flow is absorbed by the buses through some level of power demand. After the attacker has disabled some transmission lines, the physical characteristics of the generators are such that the power fed to the network can only be decreased or increased by a limited amount (called respectively \emph{ramping-down} or \emph{ramping-up} of the generator). The problem is therefore somewhat similar to a fortification problem based on a network flow, but with multiple sources and sinks and a more constrained behaviour of the base network flow problem coming from the physics of electric power generation and transmission. The power grid optimization model requires the following parameters:
\begin{itemize}
\item $P_{j,0}^g$: power output of generator $j\in J$ before any attack;
\item $RD_j$ and $RU_j$: ramping-down and ramping-up of generator $j\in J$;
\item $\bar{P}_j^g$: maximum capacity of generator $j\in J$;
\item $P_n^d$: total load at bus $n\in N$ (power demand);
\item $\bar{P}_l^f$: load flow capacity of line $l\in L$;
\item $X_l$: reactance of line $l\in L$;
\item $k_l$ and $z_l$: respectively, fortification and attack weight of line $l\in L$;
\item $K$: fortification budget;
\item $Z$: attack budget.
\end{itemize}
The model uses the following decision variables:
\begin{itemize}
\item $d_l$: binary variable equal to 1 if line $l\in L$ is fortified, 0 otherwise;
\item $a_l$: binary variable equal to 1 if line $l\in L$ is attacked, 0 otherwise;
\item $p_l^f$: power flow through transmission line $l\in L$;
\item $p_j^g$: power output of generator $j\in J$;
\item $\Delta p_n^d$: uncovered power demand at bus $n\in N$;
\item $\delta_n$: voltage phase angle (in radians) at bus $n\in N$.
\end{itemize}
A multi-level mathematical program to represent the optimisation problem can be written as follows:
\begin{subequations}
\begin{align}
\min_{\begin{subarray}{c}\\
d\in \{0,1\}^{|L|} \\[1pt]
\displaystyle \sum_{l\in L} k_l d_{l} \leq K 
 \end{subarray}} \quad 
\max_{\begin{subarray}{c}\\
a\in \{0,1\}^{|L|} \\[1pt]
\displaystyle \sum_{l\in L} z_l a_{l} \leq Z\\
\displaystyle a_l \leq 1-d_l, \forall l \in L
 \end{subarray}}
\min_{p,\Delta p,\delta} & \sum_{n\in N}\Delta p_n^d & \label{eq:obj}\\
s.t. \hspace*{0.3cm}& p_l^f = (1-a_l)\frac{1}{X_l}(\delta_{O_l}-\delta_{D_l}), & \forall l\in L \label{eq:line_power}\\
& \sum_{j\in J_n}p_j^g - \sum_{l:O_l=n}p_l^f + \sum_{l:D_l=n}p_l^f + \Delta p_n^d = P_n^d, & \forall n\in N \label{eq:flow_conservation}\\
& 0 \leq \Delta p_n^d \leq \bar P_n^d, & \forall n\in N \label{eq:unc_power}\\
& 0 \leq p_j^g \leq \bar{P}_j^g, & \forall j\in J \label{eq:gen_capacity}\\
& P_{j,0}^g-RD_j \leq p_j^g \leq P_{j,0}^g+RU_j, & j\in J \label{eq:ramping}\\
& -\bar P_l^f \leq p_l^f \leq \bar P_l^f, & \forall l\in L \label{eq:line_capacity}\\
& d_l\in\{0,1\},\ a_l\in\{0,1\}, & \forall l\in L.
\end{align}
\end{subequations}
The objective of the defender \eqref{eq:obj} is to minimise the sum of the uncovered power demands $\Delta p_n^d$ over all buses, while the attacker tries to maximise it. Both the fortification and the attack efforts are limited by their respective budgets, and attacks are restricted to non-fortified lines (see the constraints below the first $\min$ and the $\max$).
The power flowing through a transmission line is 0 in case of a line attack (constraints \eqref{eq:line_power}); otherwise, it is approximately proportional to the inverse of the line reactance and to the difference of the voltage angles at the origin and destination node. This estimate is based on considering the resistance of the transmission line as negligible and stopping the Taylor expansion of the sine of the phase difference between buses at the first order ($\sin\alpha\simeq\alpha$ if $\alpha\ll 1$). Both simplifications are generally justified when studying electric power grid interdiction (see, e.g., \cite{Salmeron2004,Wu2017}). 

After the attack on the lines, 
the demand at each bus is not necessarily covered: the classical flow conservation constraints at nodes
(constraints \eqref{eq:flow_conservation}) account for a quantity $\Delta p_n^d$ of uncovered power at each bus $n\in N$. This cannot exceed the total demand of the bus (constraints \eqref{eq:unc_power}). However, the power delivered by the generators can be adjusted only in a certain range around its former value (constraints \eqref{eq:ramping}). The power flowing through a transmission line is also bounded by the line capacity (constraints \eqref{eq:line_capacity}). 

While in classic network flow problems the demand at the sink nodes must be fully satisfied, here we try to minimise the uncovered demand.

\paragraph{\Stwo-completeness of the TEPGF problem}
We now proceed to demonstrate the completeness of the TEPGFP with unit fortification weights in the second level of the polynomial hierarchy.
\vspace{0.3cm}
\boxxx{
\textsc{\textbf{Tri-level Electric Power Grid Fortification with Unit fortification weights (TEPGFU)}}: \\
{\sc instance}: A set of generators $J$, of buses $N$, of transmission lines $L$, with a capacity and ramping-up and down parameters associated with each generator, a capacity, a reactance, an attack weight, an origin and a destination bus for each line, two positive integer budgets $K$ and $Z$ for the attack and the fortification and a positive power loss goal $\bar{K'}$.\\
{\sc question}: Is there a subset $F\subseteq L$ of lines to fortify, with $|F| \leq K$, such that for every subset of lines $A\subseteq L\setminus F$, with $\sum_{l\in A}z_l \leq Z$, the total uncovered power demand after the lines in $A$ are disabled is less than or equal to $\bar{K'}$?
}
\vspace{0.3cm}
In order to prove the \Stwo-completeness of the above problem, we will use UBIK, proved to be \Stwo-complete in Section \ref{sec:Knapsack}.

\begin{thm}
TEPGFU is \Stwo-complete.
\end{thm}
\begin{proof}{Proof}
The statement of TEPGFU is of the form $\exists F \ \ \forall A \ \ Q(F,A)$ where $Q$ can be tested in polyomial time since the optimal power flow problem is a linear program and therefore in P, directly implying that TEPGFU is in \Stwo.

We propose the following reduction from UBIK (see the obtained power grid in Figure~\ref{Fig:PowerGrid_Gadget}):
\begin{itemize}
\item We introduce a bus $b_i$ for each item $i\in I$ and an additional bus $g$; a transmission line $l_i$ goes from $g$ to $b_i$.
\item Bus $g$ is connected to a generator $j_g$ with a base power output $P^g_{j_g,0} = \sum_{i\in I} p_i$, maximum capacity $\bar P^g_{j_g}=\sum_{i\in I} p_i$, ramping-up and down $RU_{j_g} = 0$ and $RD_{j_g} = \bar P^g_{j_g}$. This implies that the generator can decrease the power flow by any amount, but not increase it.
\item The power demand of each bus $b_i$ is $P^d_{b_i}=p_i$, and the load capacity of the corresponding transmission line $l_i$ is $\bar P^f_{l_i}=p_i$. The attack weight is $z_{l_i} = w_i$, while the reactance of all lines is set at a constant $X'$ such that its inverse is much larger than any line capacity (for example, $X'\gg 1/\sum_i p_i$.
\item We set the fortification budget $K = B_2$, the attack budget $Z = W$ and the total uncovered demand goal $\bar{K'}=\bar K-1$.
\end{itemize}

\begin{figure}[ht]
\begin{centering}
\begin{tikzpicture}[>=latex, semithick]

\node [circle, draw, semithick,  minimum size=1cm, label={[label distance=0.2cm]0:$j_g$}] (jg) at (6,0) {\LARGE\AC};
\node [circle, draw=white, semithick,  minimum size=0.01cm, text width=0pt, text height=0pt, inner sep=0pt] (g) at (6,2) {};
\node [circle, draw=white, semithick,  minimum size=0.01cm, text width=0pt, text height=0pt, inner sep=0pt] (g1) at (0,2) {};
\node [circle, draw=white, semithick,  minimum size=0.01cm, text width=0pt, text height=0pt, inner sep=0pt, label={[label distance=0.1cm]0:$g$}] (g2) at (12,2) {};
\node [circle, draw=white, semithick,  minimum size=0.01cm, text width=0pt, text height=0pt, inner sep=0pt] (l11) at (0.5,2) {};
\node [circle, draw=white, semithick,  minimum size=0.01cm, text width=0pt, text height=0pt, inner sep=0pt] (l12) at (0.5,4) {};
\node [circle, draw=white, semithick,  minimum size=0.01cm, text width=0pt, text height=0pt, inner sep=0pt] (l21) at (3.5,2) {};
\node [circle, draw=white, semithick,  minimum size=0.01cm, text width=0pt, text height=0pt, inner sep=0pt] (l22) at (3.5,4) {};
\node [circle, draw=white, semithick,  minimum size=0.01cm, text width=0pt, text height=0pt, inner sep=0pt] (ln1) at (11.5,2) {};
\node [circle, draw=white, semithick,  minimum size=0.01cm, text width=0pt, text height=0pt, inner sep=0pt] (ln2) at (11.5,4) {};
\node [circle, fill=black, draw, semithick,  minimum size=0.1cm, text width=0pt, text height=0pt, inner sep=0pt] (point1) at (6.5,4) {};
\node [circle, fill=black, draw, semithick,  minimum size=0.1cm, text width=0pt, text height=0pt, inner sep=0pt] (point2) at (7.5,4) {};
\node [circle, fill=black, draw, semithick,  minimum size=0.1cm, text width=0pt, text height=0pt, inner sep=0pt] (point3) at (8.5,4) {};

\node [circle, draw=white, semithick,  minimum size=0.01cm, text width=0pt, text height=0pt, inner sep=0pt, label={[label distance=0.08cm]180:$b_1$}] (b11) at (-0.5,4) {};
\node [circle, draw=white, semithick,  minimum size=0.01cm, text width=0pt, text height=0pt, inner sep=0pt] (b12) at (1.5,4) {};
\node [circle, draw=white, semithick,  minimum size=0.01cm, text width=0pt, text height=0pt, inner sep=0pt, label={[label distance=0.08cm]180:$b_2$}] (b21) at (2.5,4) {};
\node [circle, draw=white, semithick,  minimum size=0.01cm, text width=0pt, text height=0pt, inner sep=0pt] (b22) at (4.5,4) {};
\node [circle, draw=white, semithick,  minimum size=0.01cm, text width=0pt, text height=0pt, inner sep=0pt, label={[label distance=0.08cm]180:$b_{\lvert I\rvert}$}] (bn1) at (10.5,4) {};
\node [circle, draw=white, semithick,  minimum size=0.01cm, text width=0pt, text height=0pt, inner sep=0pt] (bn2) at (12.5,4) {};

\node [circle, draw=white, semithick,  minimum size=0.01cm, text width=0pt, text height=0pt, inner sep=0pt] (d11) at (-0.3,4) {};
\node [circle, draw=white, semithick,  minimum size=0.01cm, text width=0pt, text height=0pt, inner sep=0pt] (d12) at (-0.3,5) {};
\node [circle, draw=white, semithick,  minimum size=0.01cm, text width=0pt, text height=0pt, inner sep=0pt] (d21) at (2.7,4) {};
\node [circle, draw=white, semithick,  minimum size=0.01cm, text width=0pt, text height=0pt, inner sep=0pt] (d22) at (2.7,5) {};
\node [circle, draw=white, semithick,  minimum size=0.01cm, text width=0pt, text height=0pt, inner sep=0pt] (dn1) at (10.7,4) {};
\node [circle, draw=white, semithick,  minimum size=0.01cm, text width=0pt, text height=0pt, inner sep=0pt] (dn2) at (10.7,5) {};

\draw [-, semithick](l11) -- node[left] {$l_1$} (l12);
\draw [-, semithick](l21) -- node[left] {$l_2$} (l22);
\draw [-, semithick](ln1) -- node[left] {$l_{\lvert I\rvert}$} (ln2);
\draw [-, semithick](jg) -- (g);
\draw [-, ultra thick](g1) -- (g2);
\draw [-, ultra thick](b11) -- (b12);
\draw [-, ultra thick](b21) -- (b22);
\draw [-, ultra thick](bn1) -- (bn2);
\draw [->, line width=0.7mm](d11) -- node[right] {$p_1$} (d12);
\draw [->, line width=0.7mm](d21) -- node[right] {$p_2$} (d22);
\draw [->, line width=0.7mm](dn1) -- node[right] {$p_{\lvert I\rvert}$} (dn2);

\end{tikzpicture} 
\end{centering}
\caption{Illustration of the reduction from UBIK to TEPGFU.}
\label{Fig:PowerGrid_Gadget}
\end{figure}

Notice that, as in SPIPUF, the defender and the attacker in TEPGFU
play reversed roles with respect to UBIK: fortifying lines in TEPGFU corresponds to interdicting items in UBIK, and attacking lines in SPIPUF to selecting items in UBIK.

Suppose that UBIK is a \emph{Yes} instance. For each item $i\in I$ interdicted in UBIK, we fortify the corresponding line $l_i$ in TEPGFU. Therefore, the attacker in TEPGFU can only disable the remaining lines. By construction, as in UBIK where no selection of uninterdicted items can reach the profit goal $\bar K$ respecting the capacity, in TEPGFU no subset of lines with total attack weight respecting the budget $W=Z$ can isolate from generator $j_g$ (i.e., disconnect from $j_g$) a total power demand of $\bar K$. Hence, the total uncovered demand is at most $\bar{K}-1 = \bar{K'}$. The defender can achieve this result sending a power load $p^f_{l_i}=p_i$ on each uninterdicted line, and setting the value of the voltage phase angle at the corresponding bus $b_i$ to $\delta_{b_i}=\delta_g-X'p_i$ and $\delta_g$ to an arbitrary reference value. Since there is a single generator overall and a single transmission line for each bus, from $g$ to $b_{i}$, equation~\eqref{eq:flow_conservation} applied to bus $b_i$ reduces to $0 - \sum_{l:O_l=b_{i}}p_l^f + \sum_{l:D_l=b_{i}}p_l^f + \Delta p_{b_{i}}^d = P_{b_{i}}^d$, that is $p_{l_{i}}^f + \Delta p_{b_{i}}^d = P_{b_{i}}^d$. For the buses whose line is interdicted, this further reduces to: $\Delta p_{b_i}^d=P_{b_i}^d=p_i$, and the whole power demand is uncovered. For the remaining buses, it becomes: $p_{l_i}^f+\Delta p_{b_i}^d=(\delta_g-\delta_{b_i})/X'+\Delta p_{b_i}^d=p_i+\Delta p_{b_i}^d=p_i$, therefore $\Delta p_{b_i}^d=0$, meaning that we can cover the whole demand of bus $b_i$. The total power ramping down of the generator is at most $\bar{K'}\leq RD_{j_g}$
. TEPGFU is therefore a \emph{Yes} instance.

Conversely, suppose that TEPGFU is a \emph{Yes} instance. For each fortified line in TEPGFU, we interdict the corresponding item in UBIK. For any attack in TEPGFU which preserves the set of fortified lines, the maximum total uncovered power demand is $\bar{K'}=\bar K-1$, meaning that for any set of lines with total attack weight less than or equal to $Z=W$, the corresponding total demand of the associated buses is strictly less than $\bar K$ (with voltage phase angle values similar to the ones provided in the first part of the proof). We can map any follower's solution in UBIK to an attacker's solution in TEPGFU by attacking all transmission lines associated to items which are selected by the follower. Therefore, in UBIK, the follower can never select a set of items respecting the capacity $W$ with a total profit reaching $\bar K$. UBIK is therefore a \emph{Yes} instance, which completes the proof.
\end{proof}

We observe that, as for the SPIPF, the literature on this problem usually tackles, for the sake of simplicity, the case with both unit fortification and attack weights~\cite{Chen2023,Fakhry2022,Wu2017}, though approaches with general attack weights, like \cite{Salmeron2004}, exist. Our reduction from UBIK, however, requires non-unitary attack weights to prove \Stwo-completeness.
\end{appendices}

\end{document}